\documentclass{tMPH2e}

\usepackage{verbatim}
\usepackage{graphicx}
\usepackage{wrapfig}
\usepackage{array}
\usepackage{color}

\newcommand{\cm}{cm$^{-1}$}

\newcommand{\3}{$_{3}$}

\newcommand{\p}{^\prime}
\newcommand{\pp}{^{\prime\prime}}

\begin{document}
\doi{10.1080/0026897YYxxxxxxxx}
\jvol{00}
\jnum{00} \jyear{2014} 



\title{Hybrid variation-perturbation method for calculating
rovibrational energy levels of polyatomic molecules}
\author {A.I. Pavlyuchko$^{a,b}$, S.N. Yurchenko$^b$, Jonathan Tennyson$^b$
\thanks{$^\ast$Corresponding author. Email: j.tennyson@ucl.ac.uk}
\\\vspace{6pt}
$^a${\em{ Moscow State University of Civil Engineering (MGSU), Russia}};
$^b${\em{Department of Physics and Astronomy, University College London, London, WC1E 6BT, UK}}
\\\vspace{6pt}\received{xxx July 2014}}

\maketitle

\begin{abstract}

  A procedure for calculation of rotation-vibration states of medium
  sized molecules is presented. It combines the advantages of
  variational calculations and perturbation theory. The vibrational
  problem is solved by diagonalizing a Hamiltonian matrix, which is
  partitioned into two sub-blocks. The first, smaller sub-block
  includes matrix elements with the largest contribution to the energy
  levels targeted in the calculations. The second, larger sub-block
  comprises those basis states which have little effect on these
  energy levels. Numerical perturbation theory, implemented as a
  Jacobi rotation, is used to compute the contributions from the
  matrix elements of the second sub-block. Only the first sub-block
  needs to be stored in memory and diagonalized. Calculations of the
  vibrational-rotational energy levels also employ a partitioning of
  the Hamiltonian matrix into sub-blocks, each of which corresponds
  either to a single vibrational state or a set of resonating
  vibrational states, with all associated rotational levels.
  Physically, this partitioning is efficient when the Coriolis
  coupling between different vibrational states is small. Numerical
  perturbation theory is used to include the cross-contributions from
  different vibrational states. Separate individual sub-blocks are
  then diagonalized, replacing the diagonalization of a large
  Hamiltonian matrix with a number of small matrix diagonalizations.
  Numerical examples show that the proposed hybrid
  variational-perturbation method greatly speeds up the variational
  procedure without significant loss of precision for both
  vibrational-rotational energy levels and transition intensities. The
  hybrid scheme can be used for accurate nuclear motion calculations on
  molecules with up to 15 atoms on currently available
  computers.\bigskip

\bigskip

\noindent {\bf{Acknowledgment}}
This work was supported by the ERC under Advanced Investigator Project 267219.
\end{abstract}

\section{Introduction}
\label{s:intro}

Approaches used to compute vibration-rotation energy levels and
wave functions include methods based on perturbation theory, effective
Hamiltonians and on the use of the variational principle. Each of
these methods has its advantages and disadvantages when used for
larger molecules such as those with more than ten atoms.

Historically, perturbation theory was the first method employed to treat the
many-body anharmonic problem
\cite{29Vaxxxx.PT,34Joxxxx.PT,39ScNeTh.PT,51Nixxxx.PT,58Blxxxx.PT,71AmNiTa.method,74JoPexx.PT,1976Ca.method,80TyPexx.PT},
see also the review by Klein \cite{74Klxxxx.PT}. Second-order perturbation theory is
the most common analytic treatment for estimating the vibrational energy levels
and including contributions from terms in the Hamiltonian beyond the harmonic
approximation (see, for example, Refs.
\cite{51Nixxxx.PT,61Mixxxx.PT,65Mixxxx.PT,05Baxxxx.PT}) as well as infrared
intensities \cite{10BaJuCi.method}. However, there are drawbacks inherent in
this method that prevent extensive use of it in practice. In particular the
results obtained depend on the specific form of the Hamiltonian and the
distribution of degenerate oscillators. Polyatomic molecules often show
quasi-degeneracies between vibrational energy levels and use of second-order
perturbation theory can result in significant errors. Furthermore, in practice
anharmonic effects in polyatomic molecules are sufficiently large that they are
often not converged with a second-order treatment. As a result perturbation
theory calculations usually overestimate the anharmonic corrections, sometimes
by a factor of two; see the final columns of Tables~\ref{tab:en3} and
\ref{tab:en4} given below.

Variational methods for the calculation of anharmonic energy levels
were developed independently by a number of authors
\cite{71Grxxxx.method, 71Suxxxx.method, 74BuHaBo.method,
  74BuHaxx.method, 75WhHaxx.method, 75CaKexx.method, 76WhHaxx.method,
  76CaPoxx.method, 77CaLaCu.method, 77CaPoxx.method, 80CaGiRa.method,
  80CaPoxx.method, 82CaHaxx.method, jt14, jt21, 83CaHaSu.method,
  jt18,84CrCaxx.method,jt26}.  Early implementations in general
computer codes focused on the use of basis functions for triatomic
molecules \cite{jt20,jt48,jt79,jt78}. More
recent developments have involved the increasing use of the discrete
variable representation \cite{85LiHaLi,bl89, lc00} and the extension of
the work to polyatomic molecules \cite{jt339, 07YuThJe.method,
  79Boxxxx.method, 80BoHaSa.method, 82Boxxxx.method, 83Boxxxx.method,
  83BoSexx.method, ap1, ap18, ap20, ap34, ap48, ap49, ap50, ap64,
  ap84, ap107}. The main advantage of this method is that it
allows the almost exact calculations of the vibrational-rotational energy levels and
wave functions for a given anharmonic potential \cite{jt512}.  However, variational
methods work much better for few-atom systems since the size of the basis set grows
factorially with the number of vibrational degrees of freedom in the molecule; hence the
size of the Hamiltonian matrices, which must be computed and diagonalized, also grows very
rapidly.  For this reason, variational calculations for polyatomic molecules are
difficult even on modern computers and therefore are not routinely used for molecules with
more than six atoms, when only small basis sets can be afforded. The computational errors
associated with incomplete basis sets are normally significantly larger than the errors
from perturbation approaches; this point will be discussed further in the following section.

The treatment of ro-vibration states adds extra complexity to the calculation. In this
case a rotational basis set, usually described with analytical rigid-rotor functions,
must be introduced. For high rotational quantum number, $J$, this leads to large
Hamiltonian matrices even for small polyatomic molecules.  However a two-step variational
procedure can be used to mitigate the effects of this problem \cite{jt46} meaning that it
has long been possible to compute rotational excitation up to dissociation for small
molecules \cite{jt64,jt230}.

There are a number of modifications of the variational method which
facilitates (ro-)vibrational calculations on larger molecules. For
example, the use of vibrational self-consistent field theory
by Gerber, Ratner and others \cite{79GeRa,86Bowman,88GeRa}.
Bowman, Carter and Handy
\cite{98CaBoHa.methods,04HaCaxx.method} used vibrational configuration
interaction (VCI) to reduce the size of the diagonalization for large
molecules.  If the molecule has separable degrees of freedom, a
significant reduction in the time taken to diagonalize the Hamiltonian
matrices can be obtained.  Scribano and Benoit
\cite{08ScBexx.method,10ScLaBe.method} used a hybrid approach where a
modification of the VCI method  take into account the
interactions of individual configurations using perturbation theory.
Similarly the general code MULTIMODE \cite{03BoCaHa.methods} allows
such treatments of larger systems \cite{02SePaCa.C6H6} with
approximations involving the degree of coupling between vibrational
modes.

Here we propose a new hybrid variational-perturbation theory method based on a
physically reasonable division of the large, full, variational Hamiltonian
matrix into weakly interacting sub-blocks. Second-order perturbation theory is
used to include cross-interaction effects between these sub-blocks, which are
then diagonalized separately. This allows one to replace a single
diagonalization of a large Hamiltonian matrix by a series of diagonalizations
of much smaller sub-matrices. We show that our hybrid method can greatly
accelerate the variational procedure by eliminating diagonalization of large
matrices without significant loss of precision in the computed
vibrational-rotational energy levels and the intensity of transitions between
them. This hybrid scheme is able to perform calculations for large molecules
containing up to 15 atoms on currently available computers. Finally we note
that a different version of the hybrid variational-perturbation approach has
recently been proposed by Fabri~{\it et al} \cite{14FaFuCs.method} and
implemented in the general Eckart-Watson Hamiltonian code DEWE
\cite{Matyus-DEWE-2007}. Their hybrid scheme uses a traditional single-state
ro-vibrational perturbation method based on the variationally computed
vibrational ($J=0$) eigenfunctions as a zero-order solution, where the
vibrational problem is solved variationally and the ro-vibrational energies are
derived using the second order perturbation theory expressions. This is
different from our two-step hybrid scheme, where the perturbation method is an
integral part of calculations.  We believe that this is the key to extending
variational calculations to larger molecules.

The method we propose is not dependent on the precise form of the Hamiltonian.
Instead the requirement of the Hamiltonian is that it should result in a
diagonally dominant matrix representation. For semi-rigid molecules this
criterion is naturally satisfied the Eckart-Watson Hamiltonian
\cite{watsonterm} as well as number of related Hamiltonians
\cite{88GrPa.method,79Sor,trove-paper}. However the method is unlikely to
perform so well for Hamiltonians whose emphasis is not on making producing
diagonally dominant matrices such those expressed in polyspherical coordinates
\cite{09GaIuxx.method}. In this work, all calculations are performed using
ANGMOL \cite{88GrPa.method}, a variational program for calculating
ro-vibrational spectra of general  polyatomic molecules. ANGMOL uses a
Hamiltonian expressed in curvilinear internal coordinates and an Eckart
embedding which is discussed briefly in the Appendix and extensively elsewhere
\cite{88GrPa.method}.

\section{A variational method for vibrational energy levels}
\label{s:variat-vib}

To start we consider the nature of the error arising from use of the
variational method for calculating vibrational ($J = 0$) energy levels
of polyatomic molecules.  In the variational Rayleigh-Ritz procedure, the
wave function, $\psi_n$, is generally approximated by a
finite expansion
\begin{equation}
\psi_n = \sum_i C_{i}^{(n)} \varphi_i
\end{equation}
in terms of some appropriate basis functions $\varphi_i$.
Mathematically, this procedure is equivalent to finding the
eigenvalues and eigenvectors of the Hamiltonian matrix whose elements
are given by:
\begin{equation}
H_{ij} = \langle \varphi_i \arrowvert \hat H_v \arrowvert \varphi_j \rangle ~~~.
\end{equation}
Eigenvalues of this matrix give values of the vibrational energy levels, $E_n$,
and the eigenvectors, $C_{i}^{(n)}$, give the wave function, $\psi_n$, when
combined with the basis functions. For ease of use and better convergence of
the basis functions, the $\varphi_i$ are generally chosen to form a complete
orthonormal set.  Exact values of the energy levels and eigenfunctions are only
guaranteed with an infinite number of basis functions.

The accuracy of the calculated energies depend on the number of basis
functions used. As the basis set is extended, the calculated energy
levels tend (mostly smoothly and monotonically) to their theoretical
limit. The variational limit for a group of (lowest) energy levels is
reached when the number of basis functions is sufficient for the
chosen energy levels to differ from the exact value by less than some
prescribed error.  The physical meaning of the variational limit is
that basis functions representing highly excited vibrational levels
make only a small contribution to the eigenfunctions of the low-lying
levels.

The number of basis functions needed to reach the variational limit depends on
the number of energy levels to be calculated. In general, the greater the
number of energy levels calculated, the larger the number of basis functions
required. However, a good choice of basis functions also speeds the
convergence: the closer the basis functions are to desired eigenfunctions, the
fewer  of them are required to achieve the variational limit. Mathematically,
this means that the off-diagonal matrix elements ($H_{ij}$) become small, the
eigenvalues become close to the corresponding diagonal element ($H_{ii}$), and
the corresponding eigenvectors tend towards being unit vectors.

Details of our method are give in the Appendix.
We start by defining the vibrational basis as a product of one-mode functions
\begin{equation}
\varphi_i = \prod_{m=1}^{N_c} \phi_m(v_m)
\end{equation}
where $\phi_m(v_m)$ is a either a Morse or a harmonic oscillator
function, $v_m$ is the excitation number of the $m^{\rm th}$
oscillator and $N_c$ is the number of vibrational degrees of freedom.
Morse oscillator functions are commonly used for the X--H stretches
and other stretching motion whose vibrations are strongly anharmonic,
such as doubly-bonded CO. Harmonic oscillator functions are used for
skeletal bonds in molecules with small anharmonicity, for example the
CC bonds in hydrocarbons, and for all deformation (angular)
oscillations. This form of the basis set helps to reduce the
off-diagonal elements in the Hamiltonian matrix and thus has been
found to give rapid convergence to the variational limit.

In variational procedures it is often considered desirable to use a
complete, orthonormal basis. The bound states of the Morse oscillator
form a variational basis of finite length basis and so are strictly
speaking not longer complete. It is possible to develop a complete set
based on the Morse oscillators \cite{jt14}; however, our practical
computation shows that rapid convergence can be achieved with a finite
number of Morse oscillators allowing the variational limit to be
reached.

Our implementation relies on the particular structure of the
Hamiltonian matrix ordered by increasing polyad (total vibrational
excitation) number, $N_V$
\begin{equation}
\label{e:N_V}
 N_V =   \sum_ {m = 1}^{N_c} a_m v_m
\end{equation}
where $a_m$ is some integer weighting which is often roughly proportional to
the inverse of the frequency \cite{trove-paper}. For simplicity
in this work we use $a_m = 1$ for all $m$.
This gives the size of the basis set, $M^{\rm max}_B$ in terms of the maximum
polyad number, $N_V^{\rm max}$,
\begin{equation}
\label{e:N_Bmax}
M^{\rm max}_B = \frac{(N_V^{\rm max} + N_c)!}{N_V^{\rm max}! N_c!} = \prod_{i=1}^{N_c} (N_V^{\rm max} + i)/i
\end{equation}
which increases factorially with the number of vibrational degrees of freedom,
$N_c$.

\begin{figure}[t!]
\centering
\includegraphics[width=0.4\textwidth]{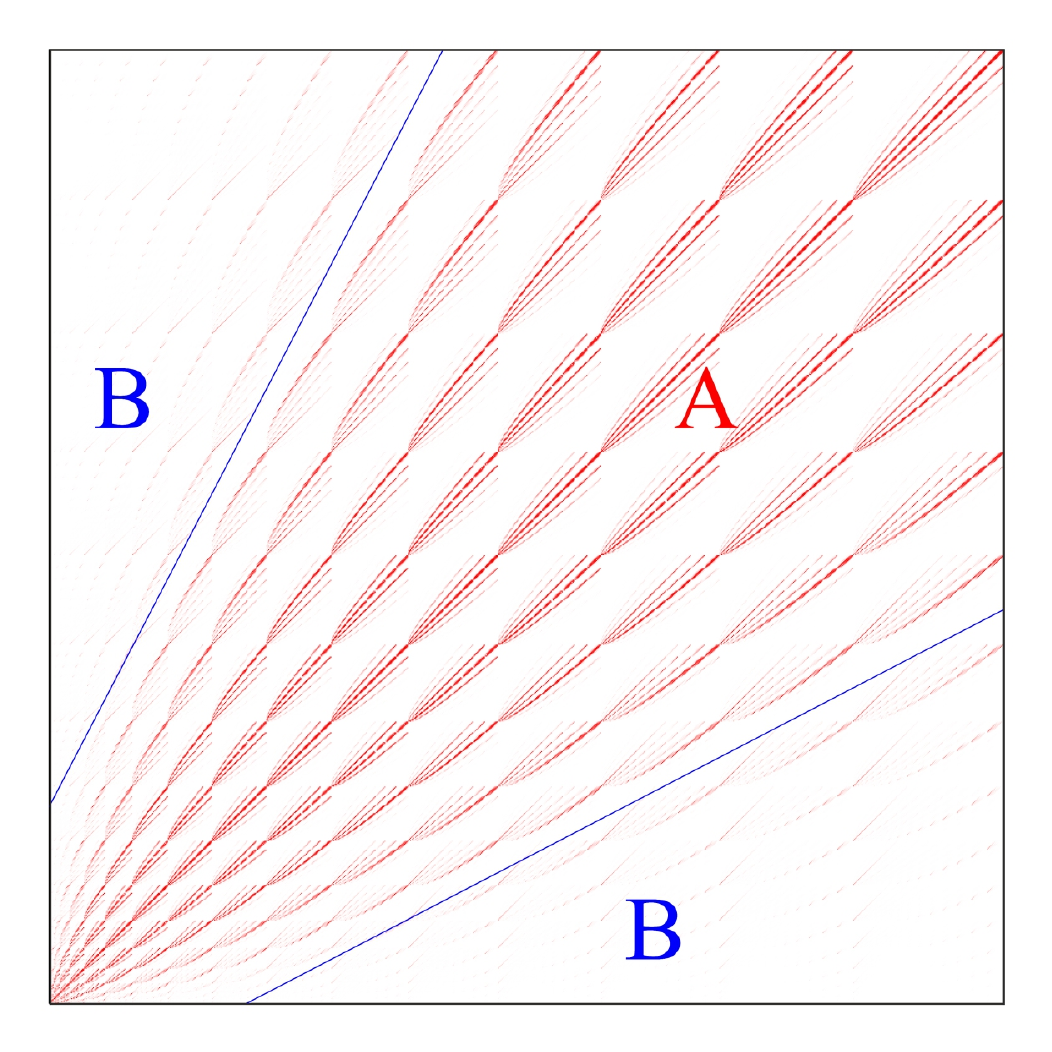}
\caption{Semi-banded structure of the vibrational ($J=0$) Hamiltonian matrix
for water ($N_V^{\rm max} = 16$,  $M^{\rm max}_B = 969$):
Region A contains large (near-)diagonal elements; Region B contains only small off-diagonal elements.
The blue continuous lines are used to separate the two regions.
The brightness of the red points is proportional to the magnitude of the individual matrix elements.}
\label{fig:1}
\end{figure}

With this definition of the basis set, the Hamiltonian matrix has a
semi-banded structure, as shown in Fig.~\ref{fig:1}, and can be
factorized into two regions: the region around the diagonal (Region
A), containing large off-diagonal elements, and the outer region
(Region B) with `small' or zero off-diagonal elements. The width of
the band, as discussed in the Appendix, depends on the form of the
Hamiltonian and obviously on the precise definition of `small'. For
example, with a potential function represented as a fourth-order
polynomial in terms of the coordinate displacements and harmonic
oscillator basis functions, the band only includes matrix elements
with excitations whose basis functions $\phi_m(v_m)$ are separated
from the diagonal by $\Delta v_m \le 4$. Use of the Morse oscillator
basis functions gives a similar picture.  Clearly distinguishable in
Fig.~\ref{fig:1} are eight off-diagonal bands corresponding to
elements connected by $\Delta v_m = \pm 1$, $\pm 2$, $\pm 3$, and $\pm
4$. Thus for a fourth-order, anharmonic potential function, all matrix
elements separated by $\Delta v_m > 4$ lie in Region~B which becomes
increasingly large as either $N_V$ or $N_c$ increases.  This rule has
similarities with the well-known Slater's rules for configuration
interaction matrix elements in electronic structure calculations.

As an example we use the program ANGMOL to compute the vibrational band origins and
corresponding band intensities for H$_2$O and HNO$_3$. In these calculations we employ
semi-empirical potential energy functions represented as fourth-order polynomials which
were obtained by fitting the corresponding potential parameters to experimentally derived
energies \cite{jt539,jt557}. These empirical potential energy functions reproduce the
experimental vibrational term values of the fundamental and first overtone states of
HNO$_3$ and H$_2$O with the root-means-squares (rms) errors less then 0.1 and 0.4
cm$^{-1}$, respectively. Initial values of the potential parameters were obtained using
{\it ab initio} calculations at the CCSD(T)/aug-cc-pVQZ level of theory. The
corresponding dipole moments were represented in the form of a second-order polynomial
fitted to the CCSD(T)/aug-cc-pVQZ {\it ab initio} values.  Full details will be given
elsewhere \cite{jtpav}.

All calculations were performed using curvilinear vibrational
coordinates, see Appendix.  To represent the stretching coordinates we
use  Morse coordinates $q_i = 1-\exp(\alpha_i \Delta r_i)$, where
$\Delta r_i$ in the bond length displacements of the $i^{\rm th}$ bond
and $\alpha_i$ is the standard Morse parameter.  Valence angles were
represented by $q_j = \cos \theta_j - \cos \theta_{\rm e}$, where
$\theta_j$ and $\theta_{\rm e}$ are the $j^{\rm th}$ inter-bond angle and its
equilibrium value, respectively.  Basis functions were constructed as
a direct product of the Morse oscillator functions for the stretches
and harmonic oscillator functions for the angles. The harmonic part of
the Hamiltonian is constructed to be diagonal for the bending part.

This form of variational basis functions has the  advantage that
the vibrational Hamiltonian matrix has relatively small off-diagonal
elements, i.e. it is close to a diagonal form.  This is due to (i) the
property of the Morse oscillators which give a good description
of the stretching
modes and (ii) small bending anharmonic terms. Besides, in this basis
the Hamiltonian matrix elements all have simple analytical forms and
thus can be efficiently evaluated.

Tables~\ref{tab:en1} and \ref{tab:en2} present calculated values of the
wavenumbers and intensities of the vibrational transitions for H$_2$O and
HNO$_3$ as a function of the number of basis functions used in the  variational
calculations. In the following, the parameter $N_{V}^{\rm target}$ ($N_{V}^{\rm
target} \le N_{V}^{\rm max}$) will be used to reference the polyad number for
energy values targeted in the variational calculations and $M_{V}^{\rm target}$
($M_{V}^{\rm target} \le M_{V}^{\rm max}$) to reference the corresponding
number of vibrational states. The results for the water molecule suggest that
to compute all vibrational  levels associated with the polyad number $N_V \le
N_{V}^{\rm target}$ to better than 5 cm$^{-1}$, the basis set must include
functions with $N_V^{\rm max}$ at least up to $N_{V}^{\rm target} + 4 $. For
example, to obtain all energies up to the second overtones ($N_{V}^{\rm target}
= 3$) with this accuracy, $N_V^{\rm max}$ must be at least 7. This is in accord
with our assumption that for the potential energy function given as a
fourth-order polynomial, the large matrix elements which belong in Region 1 are
all associated with functions given by $N_V \leq N_{V}^{\rm target} + 4$.

Calculating all $N_V \le N_{V}^{\rm target}$ vibrational term values
to an accuracy better than 0.3 cm$^{-1}$ requires basis functions
with $N_V^{\rm max} \geq N_{V}^{\rm target} +8$. This means that the
Hamiltonian matrix must include all the basis functions for which the
difference in $v_m$ is up to 8.

For water, using the lowest possible basis, {\it i.e.} $N_V^{\rm max}
= N_V^{\rm target}$, gives errors of 28 cm$^{-1}$ for
the  $\nu_2$ fundamental band, 84 cm$^{-1}$ for the first
overtone $2\nu_2$ and 154 cm$^{-1}$ for $3\nu_2$, see
Table~\ref{tab:en1}. For HNO$_3$ the situation even worse, with $N_V^{\rm max}
= N_V^{\rm target}+1$, the band centres of the fundamentals are not reproduced to
within 500 cm$^{-1}$.

It is important to note that the number of the target energy levels
defined by $N_{V}^{\rm target}$ grows rapidly with the number of
degrees of freedom $N_c$. For a triatomic molecule like water ($N_c =$ 3),
to reach an accuracy better than 0.5 cm$^{-1}$ for the fundamental
bands only ($N_{V}^{\rm target} = 1$, $N_V^{\rm max} = 9$) requires
$M_B^{\rm max} =$ 220 basis functions. For a pentatomic molecule, such as HNO$_3$
($N_c =$ 9), this requires  $M_B^{\rm max} =$ 48~620 basis functions and
for an 8-atom molecule such as C$_2$H$_6$ ($N_c$ = 18), $M_B^{\rm max}$ is 4~686~825.

The vibrational energies of polyatomic molecules are often characterized by
(accidental) resonances, such as Fermi resonances, resulting from strong
mixing between vibrationally excited states.
One consequence of this is that for HNO$_3$ $N_V^{\rm max}$
is increased to $N_V^{\rm max} \geq N_{V}^{\rm target} +9$.
Comparison of columns 2 of Tables~\ref{tab:en2} and
Tables~\ref{tab:en4} shows that even when for $M_B^{\rm max} =$ 48~620
basis functions corresponding to $N_V^{\rm max} = 9$, the convergence  of the
fundamental energy levels, {\it i.e.} $N_{V}^{\rm target}=1$, is only
good to 4 cm$^{-1}$.

For nitric acid the resonances can also give rise to large errors in
band intensities for vibrational states involved in these resonances,
especially when small basis sets are used. However, the sum of the
bands intensities for all resonant states is less sensitive to the
size of the basis set.  We note, however, that for water even individual
vibrational band intensities show relatively small dependence
on the basis set size, see Table~\ref{tab:en1}.

Tables~\ref{tab:en1} and ~\ref{tab:en2} also show that the stretching
energy levels converge faster, requiring less basis functions, than
the bending modes. This is because the Morse oscillators are more
compact than the harmonic oscillator basis functions and also because
the stretching quanta of hydrogen bonds is normally about two times
larger than that of the bending modes, i.e, it takes fewer
stretching quanta to reach the same energy. However, the
harmonic oscillator functions used for the bending modes also
represent a reasonable choice giving small coupling matrix
elements, which satisfy the requirements of perturbation theory.

\section {A hybrid method for calculating vibrational energy levels}
\label{s:hybrid-vib}

\begin{figure}[t!]
\centering
\includegraphics[width = 0.4\textwidth]{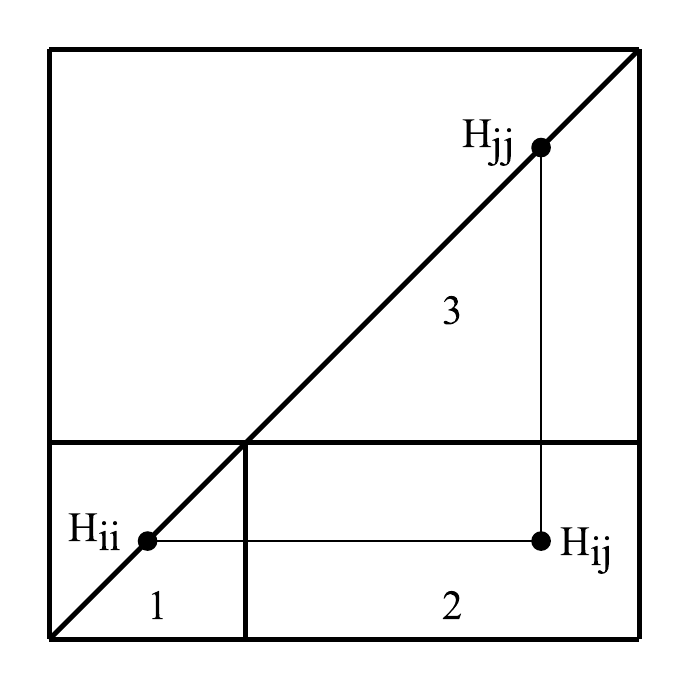}
\caption{Block structure of the vibrational Hamiltonian matrix.
    Region 1: matrix elements with the largest contributions to the
    target energy levels ($N_V \leq N_{v}^{\rm target} + 4$);
 Region 2 contains elements with small contributions to these states.
The contribution from elements in
Region 3 is disregarded.}
\label{fig:2}
\end{figure}

Variational calculations on many-atom systems rapidly become impractical. To
address this problem we implement a mixed variational-perturbation theory
approach. The idea is based on the observation that the calculated energies and
wave functions of the lower-lying vibrational levels depend differently on
different parts of the Hamiltonian matrix.  As described above, the general
Hamiltonian matrix can be factorised into two regions as illustrated in
Fig.~\ref{fig:2}. The first region contains the zeroth-order contribution to
the calculated vibrational energies as well as all other states strongly
coupled to them.  As noted above, these strongly-coupled levels are those for
which the Hamiltonian matrix contains large off-diagonal elements with the
states of interest.  For a potential function represented as a fourth-order
polynomial, the first block should contain couplings to the targeted states
with all contributions with $\Delta v_m \le 4$.

The second region of the matrix is generally much larger. This region
includes all the remaining basis states that only have a small effect
on the calculated energy levels in question. These basis states are
ones for which the polyad numbers $N_V$ differ by more than 4 from
those of the levels of interest, $N_V > N_V^{\rm target} +4$, because
the corresponding off-diagonal elements are small.

The essence of our method is that instead of diagonalizing the whole Hamiltonian matrix,
we diagonalize the smaller block 1 (see Fig.~\ref{fig:2}) with the matrix elements
corrected by the second order perturbative contributions from Region 2 as described in
detail below.

Assuming the size $M^{\rm max}_B$ of the overall  Hamiltonian matrix is defined by $N_V^{\rm max}$
in connection with Eq.~(\ref{e:N_Bmax})  and the size of block 1, $M_B^{(1)}$, is given by
\begin{equation}
\label{e:N_B}
M_B^{(1)} = \frac{(N_V^{(1)} + N_c)!}{N_V^{(1)}! N_c!} =
\prod_{i=1}^{N_c} (N_V^{(1)} + i)/i~,
\end{equation}
where $N_V^{(1)}$ is the number of the largest polyad included in block 1.  To
reach an accuracy of 1 \cm\ or better, the following two conditions have to be
met: $N_V^{\rm max} \geq N_V^{\rm target}+8$  and $N_V^{(1)} \geq N_V^{\rm
target} +4$.

In the first step of our approach all the matrix elements from blocks 1 are computed
along with the diagonal matrix elements of block 3. The second step involves computing
the off-diagonal elements of block 2 and accounting for their effect on the matrix
elements of the block 1 using one Jacobi rotation
\cite{1846Ja.method,1965Wi.method}. Considering a contribution from the block 2
off-diagonal element $H_{ij} $, which couples the diagonal elements $H_{ii}$ in block 1
and $H_{jj}$ in block 3, see Fig.~\ref{fig:2}, these two diagonal elements are
perturbatively adjusted using the Jacobi formula
\begin{equation}
\tilde{H}_{ii} = H_{ii} - \sum_{j \in {\rm block 2}} \Delta E_{ij} ~ ,
\end{equation}
where
\begin{equation}
\Delta E_{ij} = \frac{{\rm sign}(\sigma_{ij}) H_{ij} }{|\sigma_{ij}| + \sqrt{1 + \sigma_{ij}^2}}
\end{equation}
and
\begin{equation}
\sigma_{ij} = \frac{(H_{jj} - H_{ii})}{ 2 H_{ij} } ~.
\end{equation}
Here $H_{ii}$ is an initial (unperturbed) diagonal element and $\tilde{H}_{ii}$
is an adjusted one.  This formula automatically allows for cases where the
unperturbed vibrational states are (quasi-)degenerate, i.e. $H_{ii} \approx
H_{jj}$.  In the absence of the degeneracy, $|H_{ij}| \ll |H_{jj} - H_{ii}|$,
we have $\Delta E_{ij} \approx H_{ij}^{2} / (H_{jj} - H_{ ii})$, which is
equivalent to the energy correction given by second-order perturbation theory
(see e.g. \cite{34Joxxxx.PT}). Similarly, if there is degeneracy or
quasi-degeneracy $|H_{ij}| \approx |H_{jj} - H_{ii}|$, we have $\Delta E_{ij}
\leq H_{ij}$ and for degeneracy $|H_{ij}| \gg |H_{jj} - H_{ii}|$, we have
$\Delta E_{ij} \approx H_{ij}$.  That is $H_{ij}^{2} / (H_{jj} - H_{ii}) \leq
\Delta E_{ij} \leq H_{ij}$ is always satisfied.

Strictly speaking, an off-diagonal element $H_{ij}$ in block 2 should
be included as perturbative contribution both to the relevant diagonal
$H_{ii}$ and off-diagonal $H_{ik}$ elements from block 1. However, the
changes to the off-diagonal are of the second-order leading to small
contributions and are thus omitted here without significant loss of
accuracy, as will be shown below.

In the final step of the algorithm, block 1 of the corrected
Hamiltonian matrix is diagonalized. The resulting eigenfunctions are
represented as expansions in terms of the basis functions associated
with block 1 only, while the eigenvalues include contributions from
all three regions of the matrix.

As in the examples above, we use H\2O and HNO\3\ to illustrate the method (see
also Ref. \cite{06FeHaVa.PT}, where VPT2 was used to study the IR spectroscopic
properties of HNO\3). Tables \ref{tab:en3} and \ref{tab:en4} show similar
results obtained using our hybrid method, where the vibrational term values
calculated with the size of block 1, $M_B^{(1)}$, increasing. We note that when
$N_V^{(1)} = N_V^{\rm max}$, our hybrid method automatically becomes a full
variational calculation.

The hybrid method provides accurate results for the vibrational energy levels even when
the dimension of the final diagonalized block $M_B^{(1)} $ is much smaller than
the dimension of the full Hamiltonian matrix $M_B^{\rm max} $. For example, to calculate
the vibrational energy levels of water with an average accuracy of 1 cm$^{-1}$, it is
sufficient to explicitly include the basis functions with $N_V^{(1)} \geq
N_V^{\rm target} +4$. That is, block 1 includes all the larger off-diagonal elements (at
least $N_V^{\rm target}+4$) and block 2 contains the smaller off-diagonal elements
contributing to the target vibrational energies (at least $N_V^{\rm target}+8$).

This reduction is especially valuable for larger molecules. For example, the
$N_V^{(1)} = N_V^{\rm target} +4$ rule for the eight-atomic molecule C$_2$H$_6$
($N_c$ = 18) means that to calculate the fundamental band centres to within 1
cm$^{-1}$, the dimension of block 1 is only  $M_B^{(1)} = 33~649$ ($N_V^{(1)}
=$ 5), while the full size of the Hamiltonian matrix is $M_B^{\rm max} =
4~686~825$ ($N_V^{\rm max} =$ 9). For HNO$_3$ we have $M_B^{(1)} = 2~002$ and
$M_B^{\rm max} = 48~620$ ($N_V^{\rm max} =$ 9).

However in case of a strong Fermi resonance between vibrational states the accuracy
condition must be increased at least by 1 unit. For example, in order to obtain all
vibrational term values HNO$_3$ defined by the polyad number $N_V \le N_V^{\rm target}$
to within 4.0~\cm , the basis set should include at least $N_V^{(1)} \geq
N_V^{\rm target} +5$ contributions, see Table~\ref{tab:en4}.

Even with very small basis sets the hybrid method gives
reasonable results. For example, with the minimum possible size
defined by $N_V^{(1)} = N_V^{\rm target}$ the error in the band
centres of $\nu_2$, $2\nu_2$, and $3\nu_2$ of H$_2$O is 2 cm$^{-1}$, 6
cm$^{-1}$ and 15 cm$^{-1}$, respectively, which can be compared to the
error of 28 cm$^{-1}$, 84 cm$^{-1}$ and 154 cm$^{-1}$, respectively,
obtained using the purely variational procedure with the same basis
set size. This and other examples collected in Tables~\ref{tab:en3}
and \ref{tab:en4} show that our hybrid method inherits the advantage
of perturbation theory which performs reasonably well with small basis
sets.

In the limit of $M_B^{(1)} = 0$, our variation-perturbation
method turns into the pure perturbation calculation with the
perturbation treated numerically rather than analytically. For
comparison Tables~\ref{tab:en3} and \ref{tab:en4} also give results
obtained from  perturbation theory using a single Jacobi rotation for each
off-diagonal element.

Tables~\ref{tab:en3} and \ref{tab:en4} also illustrate the performance
of the hybrid approach for the vibrational band intensities of these two
molecules. These results suggest that for small basis sets the hybrid
intensities are slightly better than intensities obtained using a
direct variational treatment of the same size. This is due to the
reasonable quality achieved, even with small basis sets, of the
corresponding energy levels and eigenfunctions, see above.

It should be noted that the hybrid method requires that the
off-diagonal elements $H_{ij}$ of the block 2 are small
compared to the elements of block 1. This requirement can be
satisfied if basis functions are close to the true eigenfunctions. Our
tests show that the Morse oscillators used for the stretching
coordinates and harmonic oscillators used for the other degrees of
freedom satisfy this criterion.

Finally we note that it is not necessary to compute off-diagonal
elements in block 3.  This means that the total number of the
off-diagonal elements calculated is $M_B^{(1)} \times M_B^{\rm max}$,
which is much less than the overall size of the Hamiltonian matrix
$M_B^{\rm max} \times M_B^{\rm max}$.

\section {Hybrid method for calculating ro-vibration energy levels}
\label{s:rovib}

Inclusion of rotational motion into the problem adds an extra level of
complexity, especially when highly excited rotational states are required. The
dimension of the ro-vibrational problem $M_{B}^{\rm
  (rv)}$ is normally increased by the factor of $(2J+1)$, where $J$ is
the total angular momentum quantum number. In the variational approach
this leads to matrices which can be several order of magnitude larger
than the corresponding pure vibrational ($J=0$) Hamiltonian matrices.

In order to make the vibrational part of the basis set more compact, we
construct the ro-vibrational basis set as a direct product of the rigid-rotor
Winston's $| J,k,m \rangle $ and a set of selected vibrational
eigenfunctions of the $J=0$ problem obtained as described above. This approach
is sometimes referenced in the literature as the $J=0$ contraction
\cite{09YuBaYa.NH3}. Let us use $M_B^{ \rm vib} \ge M_B^{\rm target}$ as the
number of vibrational eigenstates for this purpose.  $M_B^{ \rm vib} $ can be
significantly reduced compared to the total number of vibrational states, $M_B^{(1)}$, from
the previous step as defined by the size of block 1. This
is because not all of the $M_B^{(1)}$ vibrational states are
equally important for the target vibrational  bands $M_B^{\rm target}$. Indeed,
because our expansion of the kinetic energy operator is truncated at the second order
(see Appendix), many matrix elements with $\Delta v > 2$  are either exactly
zero (for the Harmonic oscillators) or very small.

As an example, consider the situation where we would like to compute a spectrum
involving all ro-vibrational energies of HNO$_3$ up to $hc \cdot 4500$~\cm.
This range corresponds to $N_V^{\rm target}$ = 4 or $M_B^{\rm target}$ = 715.
From our experience of the HNO$_3$ molecule, the corresponding ro-vibrational
basis set has to include excitations at least up to $hc \cdot 6500$~\cm\ to
reach the convergence of 1~\cm, which means including about $M_B^{ \rm vib}$ =
 8000 vibrational levels. Then the corresponding rotational contribution
may include up to $J^{\rm max}=100$. This situation is common in the ExoMol
project \cite{jt528}, which computes spectra of hot molecules.  This leads to a
set of ro-vibrational Hamiltonian matrices with the dimensions $M_{B}^{\rm
(rv)}$ ranging from 8000 to $8000 \times (2J^{\rm max}+1)$ = 1~608~000.

In the following we show how to exploit the advantages of the special
structure of the ro-vibration Hamiltonian matrix in order to simplify
this problem. For large polyatomic molecules, the ro-vibrational
Hamiltonian matrix has a pronounced quasi-block character built around
vibrational states. This is due to the fact, see Appendix, that the
off-diagonal elements $H_{\lambda k,\lambda\p k\p}^{J}$ between two
different vibrational states $\lambda$ and $\lambda\p$ always
contribute substantially less to the calculated energy levels than the
off-diagonal elements within a vibrational state, $H^{J}_{\lambda
  k,\lambda k\p}$, where $k$ is the quantum number giving the
projection of $J$ on the body-fixed z-axis ($k=-J \ldots J$) and
$\lambda$ runs over the vibrational basis set ($\lambda = 1 \ldots
M_B^{ \rm vib}$). Physically, this corresponds to small Coriolis
interactions between different vibrational states, which is achieved
by the use of the Eckart embedding \cite{35Ecxxxx.method}.

\begin{figure}[t!]
\centering
\begin{minipage}[h]{0.49\linewidth}
\includegraphics[width=0.8\textwidth]{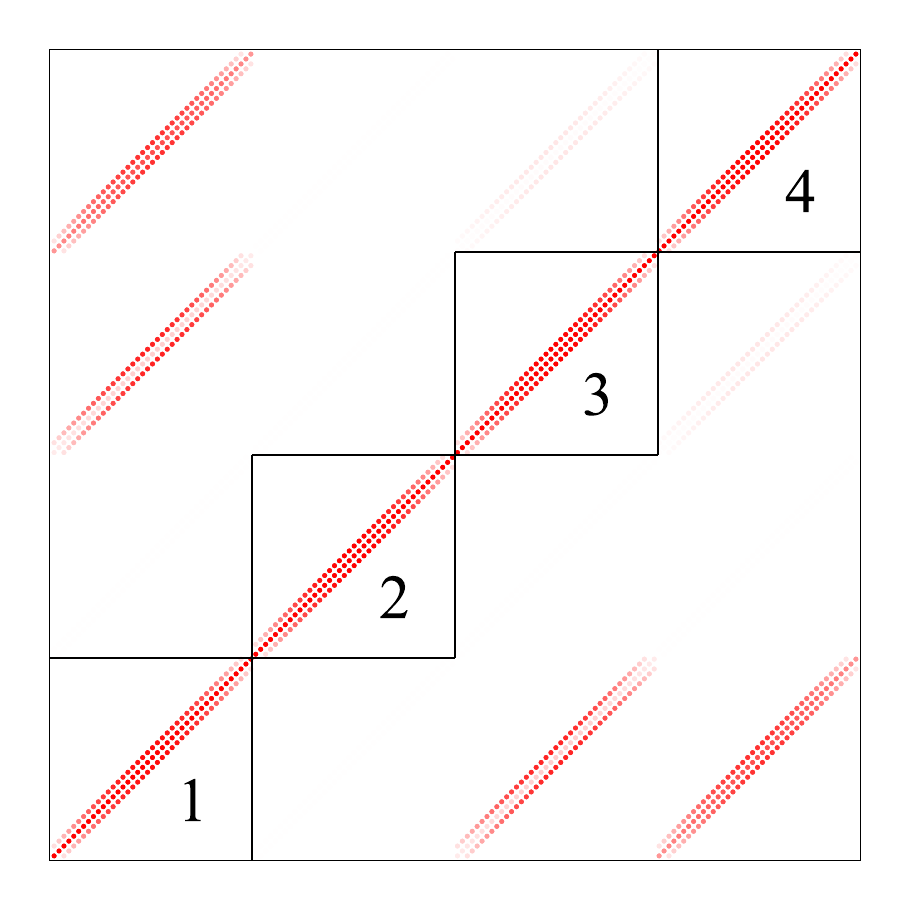} \\A
\end{minipage}
\begin{minipage}[h]{0.49\linewidth}
\includegraphics[width=0.8\textwidth]{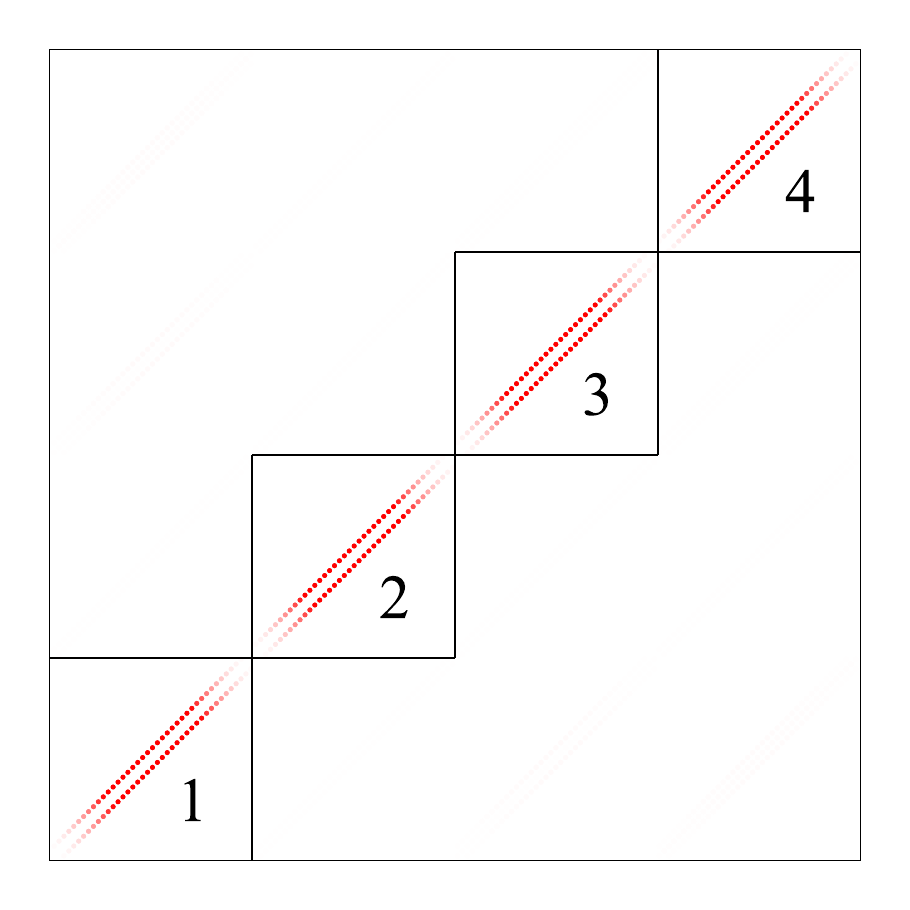} \\B
\end{minipage}
\caption{ A: Structure of the ro-vibration matrix of HNO$_3$ for the first
four vibrational states, $J=20$ (density of red is proportional to the magnitude of the individual matrix element);
B: the contributions from the off-diagonal
elements to the final energy levels (The density of red is proportional to the magnitude
of the corresponding contribution).
The squares along the diagonal depict the pure rotational sub-matrices
for vibrational states $\lambda$ = 1, 2, 3, 4 respectively.}
\label{fig:3}
\end{figure}

Figure~\ref{fig:3} illustrates the structure of the ro-vibration matrix for the
first four vibrational states of molecules HNO$_3$ ($\lambda=1\ldots 4$). We
can see that most of the off-diagonal $H^{J}_{\lambda k,\lambda\p k\p}$
elements are zero when $|k - k\p| > 2$ reflecting the quadratic differential
form of the kinetic energy operator. It can also be seen that the non-zero
off-diagonal elements gather near the main diagonal and the matrix has an
almost tridiagonal form. The near-diagonal elements $H^{J}_{\lambda k,\lambda
k\p}$ are mostly associated with changes in the effective geometry of the
rotating molecule through interactions with the corresponding vibrational state
$\lambda$.  The off-diagonal elements $H^{J}_{\lambda k,\lambda\p k\p}$ are
associated with Coriolis coupling between vibrational states $\lambda$ and
$\lambda\p$.

Importantly, the contribution of the off-diagonal elements $H^{J}_{\lambda k,\lambda\p
k\p}$ corresponding to the coupling between different vibrational states is much smaller
than that of the off-diagonal elements $H^{J}_{\lambda k,\lambda k\p}$ belonging to the
same vibrational state, because of the large energy separation between vibrational
levels. This is illustrated in Fig.~\ref{fig:3}, where the actual contributions from the
off-diagonal matrix elements to the final ro-vibrational energy levels for molecules
HNO$_3$ are shown. Typically contribution from off-diagonal elements corresponding to
different vibrational states are very small, less than 0.001~\cm.

\begin{figure}[t!]
\centering
\includegraphics[width=0.4\textwidth]{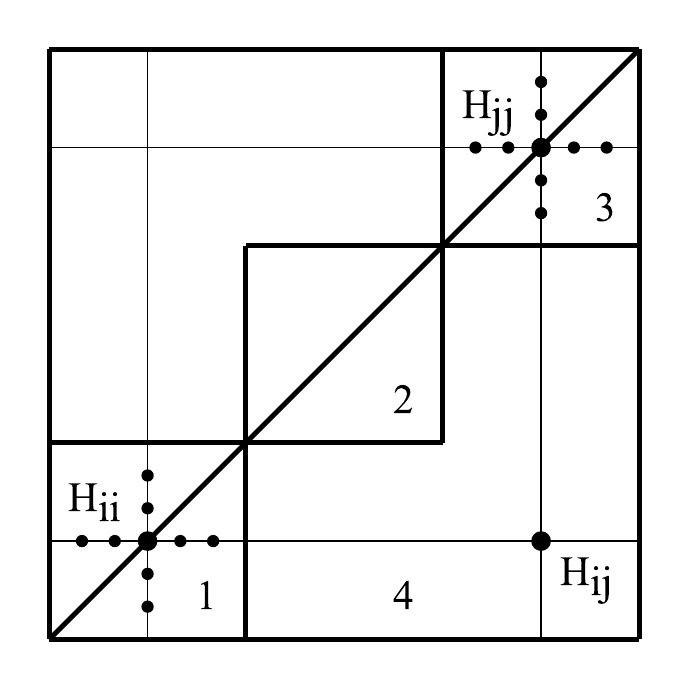}
\caption{Hybrid scheme for partitioning of ro-vibration Hamiltonian matrix:
blocks 1, 2 and 3 correspond to different vibrational states $\lambda$; the
off-diagonal matrix elements
from region 4 are generally smaller and treated using the perturbation theoryNote indices $i$ and $j$ are denoted $(\lambda k)$ and $(\lambda\p k\p)$,
respectively, in the text which distinguished between vibrational and
rotational basis functions.}
\label{fig:4}
\end{figure}

This feature of the vibration-rotation Hamiltonian matrix is common for
large molecules and underpins the ro-vibrational version of our hybrid
approach, which is schematically represented in Fig.~\ref{fig:4}. Again we use
the second-order perturbation theory (as defined by the Jacobi rotation) to
transform the $H^{J}_{\lambda k,\lambda\p k\p}$ matrix to a block diagonal form
$\tilde{H}^{J}_{\lambda k,\lambda k\p}$ built from rotational sub-matrices
corresponding to different vibrational states $\lambda$. Thus the dimension of
each degenerate rotational sub-block is $(2J+1)$ only and we only consider
$M_B^{\rm target}$  sub-matrices that correspond to  $M_B^{\rm target}$
vibrational states. In the case of strong resonances between different
vibrational states, these states can be combined into one, enlarged sub-matrix
and treated together, i.e.  for an $L$-fold symmetry vibrational degeneracy,
the dimension of the sub-block is $L \times (2J+1)$. For details see the
Appendix.

As above, we employ a single Jacobi rotation which we apply to the (complex)
ro-vibrational Hermitian matrix. Our calculations show that, in contrast to the
pure vibrational perturbation method, the best agreement with the variational
solution is achieved when both the diagonal and off-diagonal elements are
updated as given by
\begin{equation}
\tilde{H}^{J}_{\lambda k,\lambda k} = H^{J}_{\lambda k,\lambda k} + \sum_{ \lambda\p \in M_B^{\rm vib}, \lambda\p \ne \lambda} \sum_{k\p} t_{\lambda k,\lambda\p k\p} \eta_{\lambda k,\lambda\p k\p}
\end{equation}
for the diagonal elements
\begin{equation}
\label{e:H-nondiag}
\begin{split}
\tilde{H}^{J}_{\lambda k,\lambda k\pp} =
\frac{1}{2} \sum_{\lambda\p \in M_B^{\rm vib}, \lambda\p \ne \lambda} \sum_{k\p}
& \left[c_{\lambda k,\lambda\p k\p} H^{J}_{\lambda k,\lambda k\pp} + c_{\lambda k\pp,\lambda\p k\p} H^{*J}_{\lambda k,\lambda k\pp} \right. + \\
& \left. + s_{\lambda k,\lambda\p k\p} H^{J}_{\lambda\p k\p,\lambda k\pp} + H^{J}_{\lambda k,\lambda\p k\p} s_{\lambda\p k\p,\lambda k\pp} \right],
\end{split}
\end{equation}
and for the off-diagonal elements, where
\begin{eqnarray}
\nonumber
c_{\lambda k,\lambda\p k\p} &=& \frac{1}{\sqrt{1 + t_{\lambda k,\lambda\p k\p}^2}} ~, \\
\nonumber
s_{\lambda k,\lambda\p k\p} &=& \frac{c_{\lambda k,\lambda\p k\p} t_{\lambda k,\lambda\p k\p}}{\eta_{\lambda k,\lambda\p k\p}} H^{J}_{\lambda k,\lambda\p k\p} ~, \\
\nonumber
t_{\lambda k,\lambda\p k\p} &=& {\rm sign} (\vartheta_{\lambda k,\lambda\p k\p}) / (|\vartheta_{\lambda k,\lambda\p k\p}| + \sqrt{1 + \vartheta_{\lambda k,\lambda\p k\p}^2}) ~~~,~~~ \\
\nonumber
\eta_{\lambda k,\lambda\p k\p} &=& {\rm sign}\left[{\operatorname{\mathbb{R}e}( H^{J}_{\lambda k,\lambda\p k\p} ) } \right] |H^{J}_{\lambda k,\lambda\p k\p}| ~~~, \\
\nonumber
\vartheta_{\lambda k,\lambda\p k\p} &=& \frac{H^{J}_{\lambda k,\lambda k} - H^{J}_{\lambda\p k\p,\lambda\p k\p}}{2 \eta_{\lambda k,\lambda\p k\p}} ~~~.
\end{eqnarray}
Here $\underline{H}^{J}$ and $\underline{\tilde{H}}^{J}$ are the initial
(unperturbed) matrix and perturbed matrix, respectively; $\lambda$ runs from 1
to $M_B^{\rm target}$, and $k = -J \ldots +J$.

Equation~(\ref{e:H-nondiag}) is a symmetrised version of the standard
formula for the single Jacobi rotation with respect to the indices $\lambda k$
and $\lambda k\pp$. It should be noted that  in the limit $|H^{J}_{\lambda
k,\lambda\p k\p}| \ll |H^{J}_{\lambda k,\lambda k} - H^{J}_{\lambda\p
k\p,\lambda\p k\p}|$, Eq.~(\ref{e:H-nondiag}) is identical with the
corresponding expression for second-order perturbation theory:
\begin{equation}
\begin{split}
\tilde{H}^{J}_{\lambda k,\lambda k\pp} = H^{J}_{\lambda k,\lambda k\pp} +
\frac{1}{2} \sum_{\lambda\p \in M_B^{\rm vib}, \lambda\p \ne \lambda} \sum_{k\p}
& \left( \frac{H^{J}_{\lambda k,\lambda\p k\p} H^{*J}_{\lambda\p k\p,\lambda k\pp}}
{H^{J}_{\lambda k,\lambda k} - H^{J}_{\lambda\p k\p,\lambda\p k\p}} \right. + \\
& \left. + \frac{H^{J}_{\lambda k,\lambda\p k\p} H^{*J}_{\lambda\p k\p,\lambda k\pp}}
{H^{J}_{\lambda k\pp,\lambda k\pp} - H^{J}_{\lambda\p k\p,\lambda\p k\p}} \right).
\end{split}
\end{equation}

The resulting block-diagonal form is then diagonalized for each $\lambda$
sub-matrix separately, where $\lambda$ indicates either a single vibrational
state or a set of strongly interacting vibrational states. Thus our algorithm
replaces the diagonalization of a huge [$M_B^{\rm vib}\times (2J+1)$]
ro-vibrational matrix with a number of diagonalizations of much smaller
[$(2J+1)$ or $L\times (2J+1)$] matrices, where $L$ is the number of resonance
sub-states. It should be noted that in our algorithm, the ro-vibration wave
functions, in contrast to those from an exact diagonalization, do not contain
any contribution from other vibrational states.  This leads to some errors in
the subsequent calculation of the transition dipole moments and intensities of
the ro-vibrational transitions \cite{67Nixxxx.PT}. However, the relative error
is rather small for the stronger and most important transitions because of the
relatively small perturbation contribution from the off-diagonal ($\lambda \ne
\lambda\p$) matrix elements $H^{J}_{\lambda k,\lambda\p k\p}$. For other
states, where intensity stealing is important \cite{jt89}, the resonant
coupled vibrational states procedure can be used \cite{74Klxxxx.PT}. Finally,
we note that our procedure is both very quick and easily parallelized.

\section{Transition intensities}

Although the calculation of the
ro-vibration energy levels and wave functions may be long, it is only
an intermediate stage.  The final step is the calculation of
transition intensities which involves computing matrix elements of the
dipole moment functions between the computed ro-vibration wave
functions. For hot molecules, where a huge number of transition dipoles must
be evaluated, this step becomes very time consuming and even
prohibitively expensive \cite{jt564}.

Thus, even for HNO$_3$, the problem of calculating the vibration-rotation spectrum in the
range of 0--4500 cm$^{-1}$ using a standard variational method without approximations is
extremely difficult using existing computers. For larger molecules such calculations are
currently almost impossible.

The separable form of our hybrid rotation-vibration wave functions, which are obtained
in separate diagonalizations for each vibrational state, allows a
speed-up, by several orders of magnitude, of the
subsequent evaluation of transition intensities. This is because the
wave functions are much more compact and represented by a number of
independent parts. As we will show below, this approximation in case
of HNO\3\ and H\2O, leads to relatively small errors in the overall shape and
magnitude of the ro-vibrational spectra computed at room or higher
temperatures.

\section{Results}
\label{s:res}

Tables~\ref{tab:en5} and \ref{tab:en6} show the pure rotational energy
levels of H$_2$O and HNO$_3$ for several $J$ values obtained using
three different methods to find eigensolutions of the corresponding
Hamiltonian matrices: full diagionalisation (F), the hybrid
ro-vibrational approach described above (H) and a rigid rotor
calculation represented by separate diagonalisations of the rotational
sub-blocks with the off-diagonal couplings with other vibrational
states neglected (R).

The results for water, Table~\ref{tab:en5}, are of interest because
the molecule exhibits relatively large off-diagonal
elements $H^{J}_{\lambda k,\lambda\p k\p}$ and thus large centrifugal effects.

The difference $\Delta \tilde{E}^{\rm F-R} = \tilde{E}^{\rm F} - \tilde{E}^{\rm
R} $ illustrates the magnitude of the centrifugal and Coriolis contributions,
while $\Delta \tilde{E}^{\rm F-H} = \tilde{E}^{\rm F} - \tilde{E}^{\rm H} $
shows the quality of our hybrid method for different $J$. For example, for $J =
1$ the maximum value of $\Delta \tilde{E}^{\rm F-R}$ is only 0.03 cm$^{-1}$ and
$\Delta \tilde{E}^{\rm F-H}$ is within 0.005 cm$^{-1}$ and for $J=3$, $\Delta
\tilde{E}^{\rm F-R}$ is within 0.6 cm$^{-1}$ and $\Delta \tilde{E}^{\rm F-H}$
is within 0.05 cm$^{-1}$. At higher $J$ these residuals grow significantly, up
to 161 and 44~\cm, respectively, for $J=10$.

For a given $J$, both $\Delta \tilde{E}^{\rm F-R}$ and $\Delta \tilde{E}^{\rm
F-H}$ increase with increasing rotational energy. For example, for $J = 8$ for
the lowest term value of 741.25 cm$^{-1}$ ($K_A=8$), $\Delta \tilde{E}^{\rm
F-R} = 2.05$~\cm\ and $\Delta \tilde{E}^{\rm F-H}=0.02$~\cm, while for the
highest term value 1805.18 cm$^{-1}$ ($K_A=0$) we obtain $\Delta \tilde{E}^{\rm
F-R}=71.0$~\cm\ and $\Delta \tilde{E}^{\rm F-H}=13.4$~\cm.

Thus, we conclude that for a small molecule with large centrifugal
effects, like water, the hybrid method does not provide accurate
ro-vibrational energies.  Indeed, it is well known that the
$J=0$ contraction performs poorly for water \cite{lwp+89} because
of issues with linear geometry. However,
for such systems purely
variational calculations do not present a
computational problem \cite{jt378}. For larger
molecules it is generally not necessary to deal with issues
associated with quasi-linearity.

For larger molecules with smaller rotational constants, centrifugal distortion
effects are expected to decrease significantly.  Therefore we also expect the
hybrid method to perform better for such systems. Tables~\ref{tab:en6}
illustrates this effect for HNO$_3$.  For example, for $J = 60$ for the maximal
residuals we obtain $\Delta \tilde{E}^{\rm F-R}=5.63$~\cm\ and $\Delta
\tilde{E}^{\rm F-H}=1.35$~\cm, which also correspond to the highest term value
1556.46~\cm\ for the $J=60$ manifold ($K_A=0$).  For the rotational levels in
the range $1 \leq J \leq 60$, the average error of the hybrid method is within
0.02 cm$^{-1}$, i.e. much better than that for water for the same energy range.
The range $J \leq 60$ is selected as it represents the states which are
important for the room temperature ro-vibration spectrum of HNO$_3$. It should
be noted that for large molecules with small rotational constants, like HNO\3,
individual absorption lines are often not resolved because of the high density
of the lines even at room temperature. For example, the spectrum of HNO\3\
contains $2 \times 10^{10}$ lines in the region 0 -- 4000 cm$^{-1}$ with
intensities above 10$^{-27}$~cm$/$molecules at 600 K. This suggests that
calculations with the accuracy within 0.02 cm$^{-1}$ should be sufficient for
most purposes.

\begin{figure}[t!]
\centering
\begin{minipage}[h]{0.49\linewidth}
\includegraphics[width=0.8\textwidth]{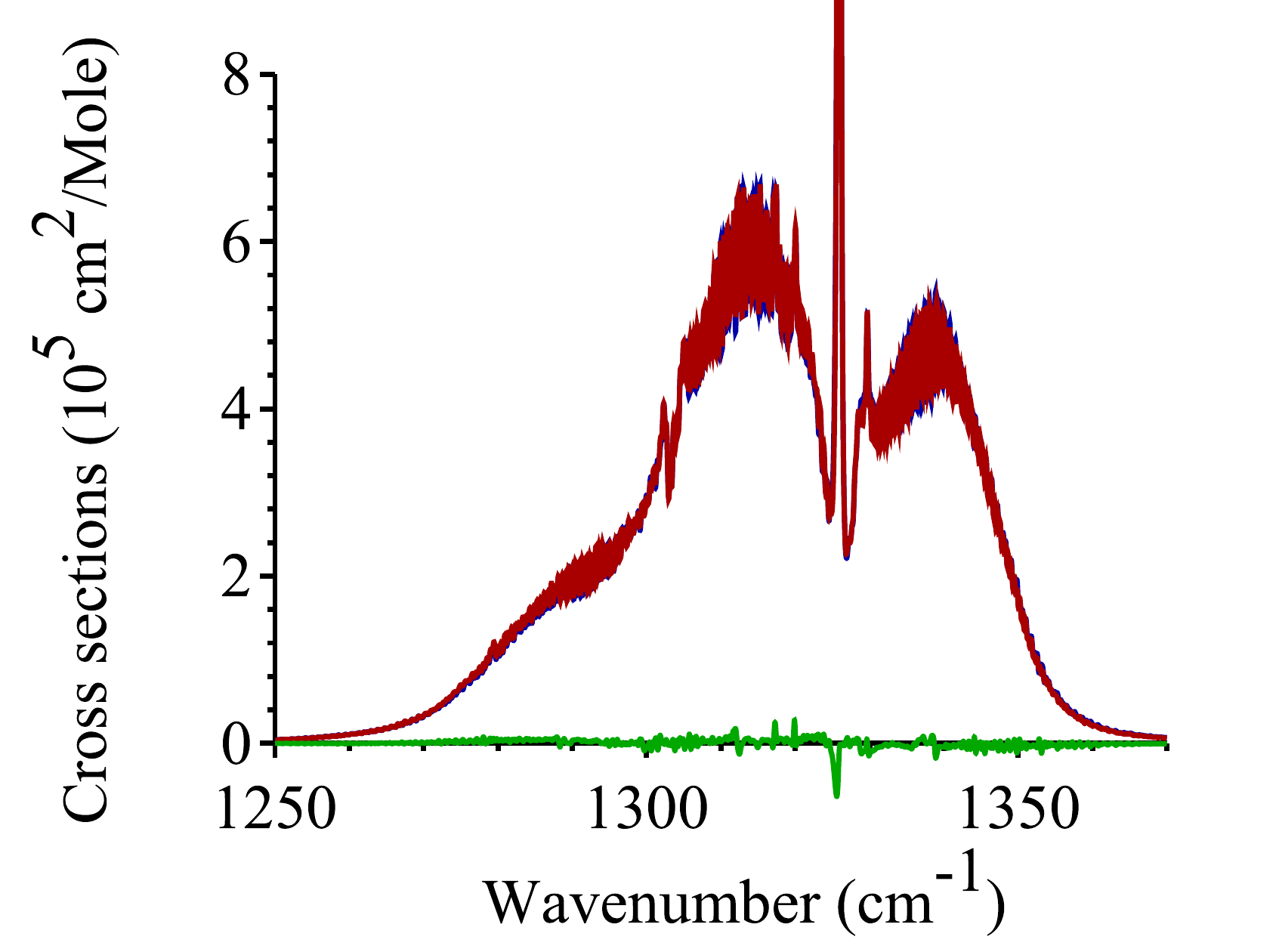} \\A
\end{minipage}
\begin{minipage}[h]{0.49\linewidth}
\includegraphics[width=0.8\textwidth]{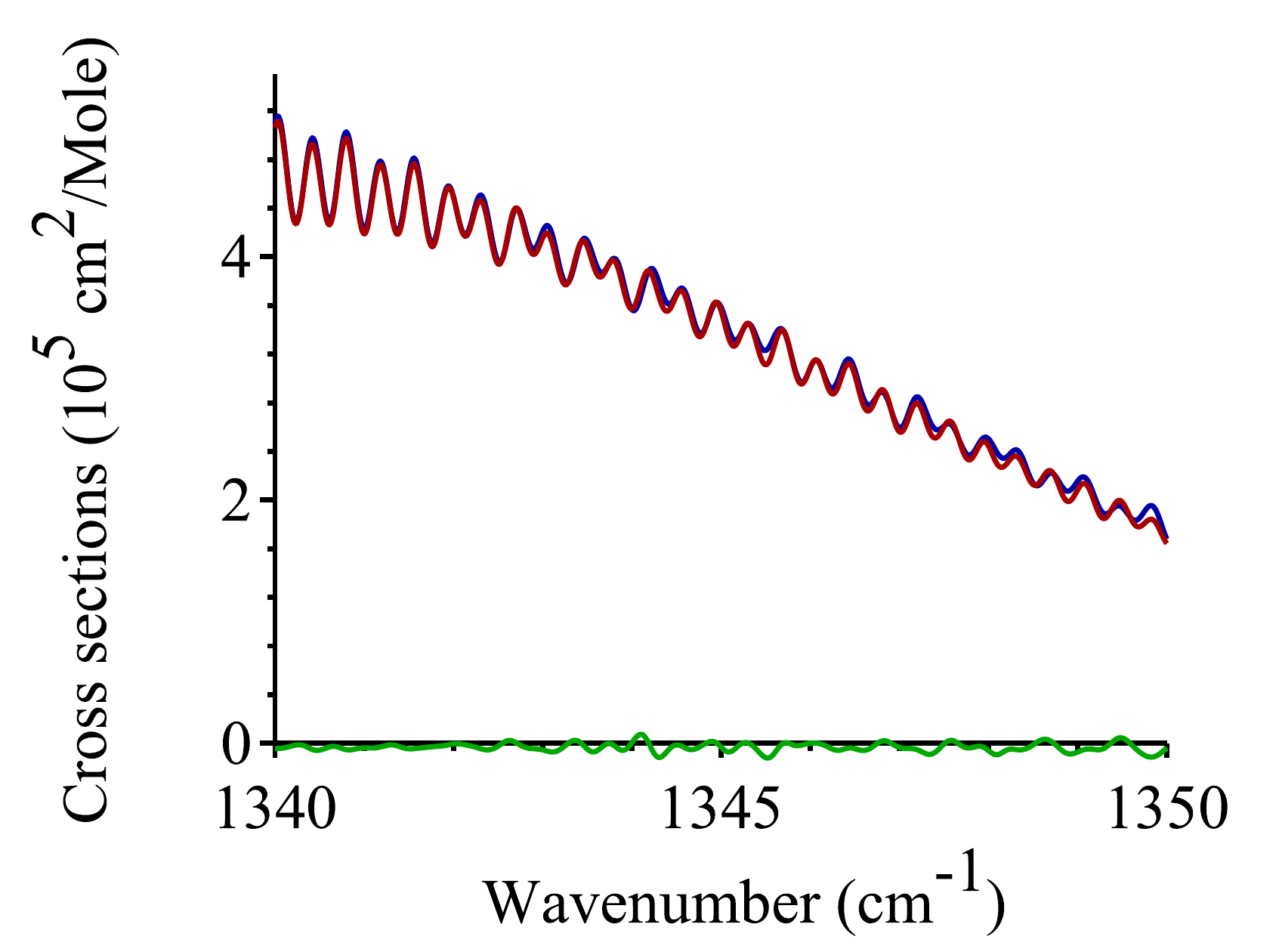} \\B
\end{minipage}
\caption{Absorption spectrum of HNO$_3$ calculated using the full diagonalization
of the ro-vibration Hamiltonian matrix (red curve), and our hybrid method (blue curve).
The green curve gives the difference between these two methods.  A: $\nu_3$ and $\nu_4$ bands;
B: $P$-branch of the $\nu_3$  band.  }
\label{fig:5}
\end{figure}

As an example, Fig.~\ref{fig:5} shows the absorption cross-sections calculated in the
region of the $\nu_3$ and $\nu_4$ bands of HNO$_3$ for $T=$ 298~K using the full
variational (`Full') calculation as well as the difference with the hybrid calculation.
Here a Voigt line profile was used with parameters $\sigma = \gamma = 0.075$ cm$^{-1}$ (a
half width at half maximum (HWHM) of 0.153 cm$^{-1}$), chosen to match spectra from the
PNNL database \cite{PNNL}. Further details of these spectra will be given elsewhere
\cite{jtpav}. For the central part of the band ($J < 40$), where the rotational lines are
strong enough to be resolved, the two methods (full and hybrid) give almost identical
results.  At the edges of the bands ($40 \le J \le 60$) slight differences in the
frequencies and intensities of the individual ro-vibration absorption lines appear.
However the rotational structure is poorly resolved due to superposition of a large
number of lines.  Besides the intensity of the band diminishes rapidly with $J$  due to
the decreasing population of the highly excited rotational levels, therefore the absolute
difference is also negligible.

\section{Conclusion}

In this paper two hybrid variation-perturbation methods for
computing vibration and ro-vibrational energies for large molecules
are proposed. These methods yield significant speed-ups for the computation
of ro-vibrational spectra, with only a minor loss of
accuracy, at least for semi-rigid polyatomic molecules. We plan to
use method to compute extensive line lists and cross sections for
large molecules, including nitric acid \cite{jtpav2}, as
part of the ExoMol project \cite{jt528}.

As an illustration of the efficiency of the hybrid method, calculations of the
vibrational and ro-vibrational energies of a tri- and a penta-atomic molecule
are presented. We can use these to project to even larger systems. Assuming
that it is currently feasible to diagonalize a matrix of about 1~000~000 by
1~000~000 and using Eq. (5), we can expect to achieve an accuracy of 1~\cm\ for
all ro-vibrational term values for polyads $N_V^{\rm} \le 4$ for a given
potential energy surface for a  molecule containing up to 15 atoms using our
hybrid procedure. For molecules containing up to 25 atoms, the expected
accuracy is about 5 cm$^{-1}$.

Our hybrid methods have several advantages:
\begin{enumerate}
\item They are relatively easy to implement as an extension to a pure
    variational computer program, which readily provides all required
    components. For example, all the
      calculations presented here were performed using ANGMOL \cite{88GrPa.method}, a
      variational program for ro-vibrational spectra of general polyatomic molecules. It
      required  changes to less than 1\%\ of the entire program to
      implement the hybrid methods.
\item They allow efficient parallelization and vectorization of the code.
\item They are very fast compared to  full diagonalizaion. Not only  is the
      diagonalization time greatly reduced, but also only a small proportion of the
      Hamiltonian matrices  needs to be computed. Furthermore, the coupling elements can
      be evaluated on the fly and do not need to be stored in the memory.
    \item Our method adopts several approximations arising from second-order
        perturbation theory and the rigid rotor model to full diagonalization
        approaching the variational limit. For the vibration-only problem,
        this approximation is easily controlled by altering the size of the
        block 1.

\end{enumerate}

 Our method has scope for further improvement. For example:
\begin{itemize}
\item For the purely vibrational Hamiltonian matrix, one can change
      not only the diagonal but also the off-diagonal elements in block 1, as
      we do for the ro-vibrational Hamiltonian matrix.  This will increase the
      accuracy at the expense of increased computer time.
\item One could use not only one but two or more consecutive
      Jacobi rotations. Again,   this will increase the
      accuracy at the expense of increased computer time.
\item Simple perturbation theory corrections generally overestimate the effect
      of the perturbation. One could reduce this effect by multiplying
      the computed perturbations by some empirical coefficient $0 < p \leq 1$.
      This will increase the accuracy without increasing the computer time.
      The problem with this approach is that the magnitude of $p$ depends on
      the number of degrees-of-freedom involved, the structure and size of
      the Hamiltonian matrix and value of perturbation adjustment.
    \item Finally, we note that in our present implementation no
      advantage is taken of the pronounced polyad structure shown by
      many molecules. This could be achieved by a suitable choice of
      the $\underline{\alpha}$ coefficients in Eq.~(\ref{e:N_V}).
\end{itemize}

\section*{Acknowledgments}
This work was supported by the ERC under Advanced Investigator Project 267219.

\clearpage

\newpage

\begin{table}[ht]
\scriptsize
\tabcolsep=3pt
\caption{Fully variational calculation of the vibrational band origins
(upper row, cm$^{-1}$) and vibrational band intensities (lower row, km / mole)
for H$_2$O as a function of total vibrational excitation number, $N^{\rm max}_V$ ,
and number of basis functions, $M^{\rm max}_B$} \label{tab:en1}
\begin{center}
\begin{tabular}{lrrrrrrrrrrrr}
\hline\hline
Transition  &  \multicolumn{12}{c}{$N^{\rm max}_V$/$M^{\rm max}_B$}  \\
\cline{2-13}
         &  20/1771 &  11/364  &  10/286  &  9/220   &  8/165   &  7/120   &  6/84    &  5/56    &  4/35    &  3/20    &  2/10    &  1/4     \\
\hline
0  1  0  &  1595.0  &  1595.0  &  1595.0  &  1595.0  &  1595.0  &  1595.1  &  1595.2  &  1595.3  &  1598.4  &  1601.6  &  1626.1  &  1623.4  \\
         &  68.72   &  68.72   &  68.72   &  68.72   &  68.73   &  68.73   &  68.73   &  68.75   &  68.85   &  68.80   &  70.11   &  69.78   \\
0  2  0  &  3152.0  &  3152.0  &  3152.1  &  3152.2  &  3152.5  &  3153.3  &  3153.5  &  3163.0  &  3173.4  &  3211.5  &  3236.0  &          \\
         &  0.424   &  0.425   &  0.425   &  0.425   &  0.426   &  0.428   &  0.432   &  0.443   &  0.502   &  0.636   &  0.773   &          \\
1  0  0  &  3656.8  &  3656.8  &  3656.8  &  3656.8  &  3656.8  &  3656.9  &  3656.9  &  3657.1  &  3657.9  &  3659.3  &  3682.0  &  3672.4  \\
         &  2.807   &  2.807   &  2.806   &  2.806   &  2.806   &  2.805   &  2.804   &  2.796   &  2.773   &  2.691   &  2.638   &  1.316   \\
0  0  1  &  3755.5  &  3755.5  &  3755.5  &  3755.5  &  3755.5  &  3755.5  &  3755.5  &  3755.6  &  3756.0  &  3756.4  &  3768.3  &  3767.4  \\
         &  49.55   &  49.55   &  49.55   &  49.55   &  49.55   &  49.55   &  49.55   &  49.56   &  49.54   &  49.53   &  49.70   &  52.41   \\
0  3  0  &  4666.8  &  4667.1  &  4667.6  &  4668.9  &  4671.2  &  4671.7  &  4694.7  &  4716.8  &  4766.5  &  4820.7  &          &          \\
         &  0.080   &  0.080   &  0.080   &  0.080   &  0.079   &  0.078   &  0.078   &  0.071   &  0.066   &  0.002   &          &          \\
1  1  0  &  5234.6  &  5234.6  &  5234.6  &  5234.7  &  5234.8  &  5235.2  &  5235.7  &  5239.7  &  5247.4  &  5297.2  &  5306.8  &          \\
         &  0.083   &  0.083   &  0.083   &  0.083   &  0.083   &  0.083   &  0.084   &  0.084   &  0.090   &  0.076   &  0.024   &          \\
0  1  1  &  5331.2  &  5331.2  &  5331.2  &  5331.2  &  5331.3  &  5331.4  &  5331.6  &  5333.8  &  5337.2  &  5365.8  &  5383.7  &          \\
         &  3.840   &  3.840   &  3.840   &  3.840   &  3.840   &  3.840   &  3.841   &  3.837   &  3.856   &  3.853   &  3.571   &          \\
1  2  0  &  6773.7  &  6773.9  &  6774.2  &  6774.8  &  6776.2  &  6777.6  &  6789.2  &  6808.7  &  6883.0  &  6914.0  &          &          \\
         &  0.016   &  0.016   &  0.016   &  0.016   &  0.017   &  0.017   &  0.018   &  0.021   &  0.037   &  0.021   &          &          \\
0  2  1  &  6870.6  &  6870.6  &  6870.7  &  6870.9  &  6871.6  &  6872.0  &  6878.8  &  6887.2  &  6935.1  &  6962.3  &          &          \\
         &  0.044   &  0.044   &  0.044   &  0.044   &  0.044   &  0.044   &  0.045   &  0.049   &  0.078   &  0.056   &          &          \\
2  0  0  &  7202.0  &  7202.0  &  7202.0  &  7202.0  &  7202.1  &  7202.3  &  7202.6  &  7203.8  &  7207.9  &  7241.0  &  7244.8  &          \\
         &  0.387   &  0.387   &  0.387   &  0.387   &  0.387   &  0.387   &  0.386   &  0.383   &  0.368   &  0.370   &  0.387   &          \\
1  0  1  &  7250.1  &  7250.1  &  7250.1  &  7250.1  &  7250.2  &  7250.3  &  7250.5  &  7251.2  &  7253.8  &  7279.2  &  7290.7  &          \\
         &  3.037   &  3.037   &  3.037   &  3.037   &  3.036   &  3.036   &  3.035   &  3.032   &  3.007   &  3.042   &  3.268   &          \\
0  0  2  &  7444.7  &  7444.7  &  7444.7  &  7444.7  &  7444.7  &  7444.8  &  7444.9  &  7445.4  &  7447.0  &  7462.7  &  7488.5  &          \\
         &  0.027   &  0.027   &  0.027   &  0.027   &  0.027   &  0.027   &  0.027   &  0.027   &  0.027   &  0.032   &  0.037   &          \\
3  0  0  & 10599.3  & 10599.4  & 10599.4  & 10599.5  & 10599.8  & 10600.4  & 10601.9  & 10611.9  & 10649.9  & 10661.9  &          &          \\
         &  0.023   &  0.023   &  0.023   &  0.023   &  0.023   &  0.023   &  0.023   &  0.021   &  0.020   &  0.031   &          &          \\
0  0  3  & 11032.7  & 11032.7  & 11032.8  & 11032.8  & 11032.9  & 11033.1  & 11033.5  & 11036.1  & 11054.7  & 11106.8  &          &          \\
         &  0.004   &  0.004   &  0.004   &  0.004   &  0.004   &  0.004   &  0.004   &  0.004   &  0.003   &  0.007   &          &          \\
\hline\hline
\end{tabular}
\end{center}
\end{table}

\begin{table}[ht]
\scriptsize
\tabcolsep=3pt
\caption{Fully variational calculation of the vibrational band origins
(upper row, cm$^{-1}$) and vibrational band intensities (lower row, km / mole)
for HNO$_3$ as a function of total vibrational excitation number, $N^{\rm max}_V$,
and number of basis functions, $M^{\rm max}_B$} \label{tab:en2}
\begin{center}
\begin{tabular}{lrrrrrrrrr}
\hline\hline
Transition  &  \multicolumn{9}{c}{$N^{\rm max}_V$/$M^{\rm max}_B$}  \\
\cline{2-10}
                   &  9/48620 &  8/24310 &  7/11440 &  6/5005  &  5/2002  &  4/715   &  3/220   &  2/55    &  1/10    \\
\hline
 $\nu_9$           &   458.8  &   459.5  &   460.4  &   466.3  &   465.2  &   509.3  &   476.1  &   945.6  &   485.5  \\
                   &   107.3  &   107.5  &   107.7  &   109.0  &   108.9  &   117.8  &   112.0  &   13.1   &   114.5  \\
 $\nu_8$           &   581.2  &   582.9  &   583.1  &   592.5  &   590.7  &   651.7  &   616.0  &  1041.3  &   688.4  \\
                   &   7.52   &   7.54   &   7.63   &   7.77   &   7.93   &   8.94   &   8.9    &   11.7   &   14.2   \\
 $\nu_7$           &   648.5  &   649.8  &   652.1  &   662.6  &   663.5  &   731.0  &   695.3  &  1315.3  &   781.8  \\
                   &   10.5   &   10.6   &   10.4   &   10.8   &   10.3   &   11.1   &   9.58   &   171.0  &   6.26   \\
 $\nu_6$           &   763.5  &   764.0  &   764.4  &   769.9  &   767.0  &   810.3  &   772.2  &   750.6  &   780.2  \\
                   &   7.90   &   7.90   &   7.90   &   7.95   &   7.91   &   8.27   &   7.99   &   158.6  &   8.05   \\
 $\nu_5$           &   881.3  &   883.1  &   885.8  &   896.3  &   899.0  &   964.1  &   934.6  &  1351.3  &  1047.1  \\
                   &   106.7  &   110.3  &   126.1  &   118.8  &   148.6  &   94.4   &   140.6  &   171.3  &   105    \\
2$\nu_9$           &   899.5  &   901.4  &   907.8  &   915.8  &   955.6  &   982.7  &  1199.7  &          &          \\
                   &   52.3   &   48.6   &   32.9   &   40.9   &   8.02   &   70.0   &   1.44   &          &          \\
 $\nu_6$ + $\nu_9$ &  1207.8  &  1209.1  &  1215.1  &  1219.7  &  1253.2  &  1272.2  &  1519.9  &          &          \\
                   &   9.69   &   9.74   &   10.5   &   10.0   &   18.8   &   9.62   &   5.29   &          &          \\
3$\nu_9$           &  1301.2  &  1310.0  &  1323.4  &  1391.8  &  1406.5  &  1671.5  &  1409.8  &          &          \\
                   &   0.266  &   0.272  &   0.263  &   0.043  &   0.181  &   0.115  &   0.007  &          &          \\
 $\nu_4$           &  1306.8  &  1297.4  &  1302.6  &  1311.3  &  1316.6  &  1368.1  &  1321.8  &  1629.2  &  1359.3  \\
                   &   55.6   &   46.5   &   74.2   &   68.1   &   78.1   &   40.3   &   52.2   &   37.3   &   42.5   \\
 $\nu_3$           &  1327.5  &  1328.9  &  1331.6  &  1343.6  &  1347.8  &  1407.4  &  1369.8  &  1744.0  &  1484.0  \\
                   &   223.7  &   227.0  &   209.6  &   163.5  &   122.5  &   256.7  &   242.2  &   173.3  &   191.9  \\
4$\nu_9$           &  1702.5  &  1723.0  &  1768.0  &  1836.2  &  2085.6  &  2178.9  &          &          &          \\
                   &   13.8   &   3.37   &   0.185  &   0.009  &   0.030  &   0.074  &          &          &          \\
 $\nu_2$           &  1710.6  &  1711.6  &  1712.7  &  1723.4  &  1722.2  &  1791.8  &  1755.0  &  2141.7  &  1875.0  \\
                   &   348.0  &   359.3  &   363.1  &   365.3  &   358.8  &   373.4  &   340.2  &   390.7  &   283.2  \\
 $\nu_1$           &  3552.0  &  3552.4  &  3553.3  &  3558.3  &  3555.8  &  3599.0  &  3559.8  &  3843.1  &  3565.5  \\
                   &   35.3   &   84.4   &   39.8   &   82.1   &   83.7   &   79.9   &   67.6   &   86.8   &   90.9   \\
\hline\hline
\end{tabular}
\end{center}
\end{table}

\begin{table}[ht]
\scriptsize
\tabcolsep=3pt
\caption{Calculated values of the vibrational band origins (upper row, cm$^{-1}$)
and vibrational band intensities (lower row, km / mole) for  H$_2$O using
our hybrid method and perturbation theory (PT)} \label{tab:en3}
\begin{center}
\begin{tabular}{crrrrrrrrrrrrr}
\hline\hline
Transition  &  \multicolumn{12}{c}{$N^{\rm max}_V$=20 , $M^{\rm max}_B$=1771 ,
$N^{(1)}_V$/$M^{(1)}_B$} & PT  \\
\cline{2-13}
         &  20/1771 &  11/364  &  10/286  &  9/220   &  8/165   &  7/120   &  6/84    &  5/56    &  4/35    &  3/20    &  2/10    &  1/4     &  \\
\hline
0  1  0  &  1595.0  &  1595.0  &  1595.0  &  1595.0  &  1594.9  &  1595.0  &  1595.0  &  1594.8  &  1594.8  &  1594.6  &  1594.7  &  1593.8  &  1593.7  \\
         &  68.72   &  68.72   &  68.72   &  68.72   &  68.72   &  68.72   &  68.72   &  68.73   &  68.69   &  68.47   &  68.72   &  68.49   &  68.58   \\
0  2  0  &  3152.0  &  3152.0  &  3152.0  &  3151.9  &  3151.9  &  3152.1  &  3151.2  &  3152.1  &  3152.8  &  3147.8  &  3146.5  &          &  3146.6  \\
         &  0.424   &  0.424   &  0.424   &  0.424   &  0.424   &  0.425   &  0.426   &  0.429   &  0.484   &  0.594   &  0.705   &          &  0.442   \\
1  0  0  &  3656.8  &  3656.8  &  3656.8  &  3656.8  &  3656.8  &  3656.8  &  3656.8  &  3656.8  &  3656.8  &  3657.0  &  3664.5  &  3660.2  &  3660.1  \\
         &  2.807   &  2.807   &  2.807   &  2.807   &  2.807   &  2.807   &  2.806   &  2.803   &  2.787   &  2.740   &  2.700   &  1.324   &  2.653   \\
0  0  1  &  3755.5  &  3755.5  &  3755.5  &  3755.5  &  3755.5  &  3755.5  &  3755.4  &  3755.4  &  3755.1  &  3754.5  &  3751.1  &  3755.3  &  3756.6  \\
         &  49.55   &  49.55   &  49.55   &  49.55   &  49.55   &  49.55   &  49.55   &  49.55   &  49.52   &  49.45   &  49.42   &  52.24   &  49.83   \\
0  3  0  &  4666.8  &  4666.6  &  4666.1  &  4666.5  &  4667.1  &  4663.6  &  4668.1  &  4672.2  &  4654.0  &  4651.7  &          &          &  4656.0  \\
         &  0.080   &  0.080   &  0.080   &  0.080   &  0.080   &  0.079   &  0.080   &  0.075   &  0.072   &  0.002   &          &          &  0.073   \\
1  1  0  &  5234.6  &  5234.6  &  5234.6  &  5234.6  &  5234.6  &  5234.7  &  5234.6  &  5234.6  &  5235.9  &  5250.4  &  5248.5  &          &  5242.7  \\
         &  0.083   &  0.083   &  0.083   &  0.083   &  0.083   &  0.083   &  0.083   &  0.084   &  0.088   &  0.082   &  0.026   &          &  0.115   \\
0  1  1  &  5331.2  &  5331.2  &  5331.2  &  5331.1  &  5331.1  &  5331.1  &  5331.0  &  5329.7  &  5327.8  &  5320.6  &  5325.6  &          &  5331.9  \\
         &  3.840   &  3.840   &  3.840   &  3.840   &  3.840   &  3.840   &  3.840   &  3.838   &  3.861   &  3.885   &  3.568   &          &  3.686   \\
1  2  0  &  6773.7  &  6773.6  &  6773.6  &  6773.7  &  6774.1  &  6773.3  &  6774.1  &  6778.9  &  6794.0  &  6802.8  &          &          &  6786.5  \\
         &  0.016   &  0.016   &  0.016   &  0.016   &  0.016   &  0.016   &  0.017   &  0.018   &  0.024   &  0.012   &          &          &  0.026   \\
0  2  1  &  6870.6  &  6870.6  &  6870.4  &  6870.4  &  6870.4  &  6869.8  &  6866.8  &  6863.8  &  6850.7  &  6850.6  &          &          &  6866.3  \\
         &  0.044   &  0.044   &  0.044   &  0.043   &  0.044   &  0.044   &  0.043   &  0.046   &  0.062   &  0.040   &          &          &  0.094   \\
2  0  0  &  7202.0  &  7202.0  &  7202.0  &  7202.0  &  7202.0  &  7202.0  &  7202.0  &  7201.8  &  7202.5  &  7214.9  &  7206.2  &          &  7202.5  \\
         &  0.387   &  0.387   &  0.387   &  0.387   &  0.387   &  0.387   &  0.387   &  0.385   &  0.377   &  0.375   &  0.386   &          &  0.369   \\
1  0  1  &  7250.1  &  7250.1  &  7250.1  &  7250.0  &  7250.0  &  7250.0  &  7250.0  &  7249.4  &  7249.0  &  7253.2  &  7252.4  &          &  7255.5  \\
         &  3.037   &  3.037   &  3.037   &  3.037   &  3.037   &  3.037   &  3.037   &  3.035   &  3.021   &  3.056   &  3.259   &          &  3.008   \\
0  0  2  &  7444.7  &  7444.7  &  7444.7  &  7444.7  &  7444.6  &  7444.6  &  7444.6  &  7444.0  &  7443.0  &  7438.5  &  7449.4  &          &  7452.2  \\
         &  0.027   &  0.027   &  0.027   &  0.027   &  0.027   &  0.027   &  0.027   &  0.027   &  0.027   &  0.033   &  0.036   &          &  0.023   \\
3  0  0  & 10599.3  & 10599.3  & 10599.2  & 10599.2  & 10599.2  & 10599.1  & 10598.4  & 10599.8  & 10613.4  & 10603.1  &          &          & 10604.9  \\
         &  0.023   &  0.023   &  0.023   &  0.023   &  0.023   &  0.023   &  0.023   &  0.023   &  0.023   &  0.031   &          &          &  0.026   \\
0  0  3  & 11032.7  & 11032.7  & 11032.7  & 11032.7  & 11032.6  & 11032.5  & 11031.5  & 11029.8  & 11022.3  & 11046.0  &          &          & 11041.2  \\
         &  0.004   &  0.004   &  0.004   &  0.004   &  0.004   &  0.004   &  0.004   &  0.004   &  0.003   &  0.007   &          &          &  0.006   \\
\hline\hline
\end{tabular}
\end{center}
\end{table}

\begin{table}[ht]
\scriptsize
\tabcolsep=3pt
\caption{Calculated values of vibrational band origins (upper row, cm$^{-1}$)
and vibrational band intensities (lower row, km / mole) for HNO$_3$ using
our hybrid method and perturbation theory (PT)}\label{tab:en4}
\begin{center}
\begin{tabular}{lrrrrrrrrrr}
\hline\hline
Transition  &  \multicolumn{9}{c}{$N^{\rm max}_V$=14 , $M^{\rm max}_B$=817190 ,
$N^{(1)}_V$/$M^{(1)}_B$} & PT  \\
\cline{2-10}
                   &  9/48620 &  8/24310 &  7/11440 &  6/5005  &  5/2002  &  4/715   &  3/220   &  2/55    &  1/10    &  \\
\hline
 $\nu_9$           &   458.2  &   457.4  &   457.7  &   453.6  &   455.0  &   448.3  &   442.0  &   486.7  &   458.5  &   458.7  \\
                   &   107.2  &   107.0  &   107.0  &   105.9  &   106.2  &   102.9  &   103.6  &   96.3   &   108.0  &   107.5  \\
 $\nu_8$           &   580.3  &   578.8  &   578.3  &   571.8  &   571.8  &   559.6  &   554.1  &   601.5  &   574.2  &   583.4  \\
                   &   7.47   &   7.39   &   7.40   &   6.99   &   6.96   &   6.40   &   6.08   &   6.88   &   11.4   &   19.8   \\
 $\nu_7$           &   646.7  &   644.2  &   644.7  &   633.8  &   635.3  &   620.4  &   610.0  &   672.9  &   635.5  &   652.5  \\
                   &   10.5   &   10.5   &   10.5   &   10.7   &   10.4   &   10.9   &   10.8   &   9.18   &   8.82   &   9.05   \\
 $\nu_6$           &   763.2  &   762.4  &   762.9  &   759.0  &   760.7  &   753.7  &   750.6  &   791.3  &   762.2  &   761.2  \\
                   &   7.90   &   7.89   &   7.88   &   7.83   &   7.81   &   7.64   &   7.72   &   6.83   &   7.84   &   8.02   \\
 $\nu_5$           &   879.1  &   877.8  &   876.4  &   870.6  &   868.6  &   856.5  &   857.0  &   918.7  &   904.5  &   899.5  \\
                   &   96.2   &   100.4  &   86.2   &   99.0   &   86.9   &   45.9   &   133.9  &   33.7   &   85.7   &   4.42   \\
2$\nu_9$           &   896.4  &   895.3  &   893.6  &   887.3  &   887.0  &   874.5  &   911.5  &   948.8  &          &   954.1  \\
                   &   62.7   &   58.3   &   73.4   &   58.5   &   71.4   &   107.5  &   8.20   &   95.7   &          &   204.8  \\
 $\nu_6$ + $\nu_9$ &  1205.4  &  1204.8  &  1200.8  &  1198.9  &  1195.4  &  1179.3  &  1212.5  &  1226.6  &          &  1212.6  \\
                   &   9.53   &   9.60   &   7.45   &   6.20   &   8.17   &   6.88   &   5.55   &   10.4   &          &   2.80   \\
3$\nu_9$           &  1288.8  &  1285.0  &  1280.7  &  1274.7  &  1274.1  &  1278.2  &  1318.7  &          &          &  1323.2  \\
                   &   0.262  &   0.250  &   0.259  &   0.273  &   0.234  &   0.130  &   0.016  &          &          &   0.032  \\
 $\nu_4$           &  1302.9  &  1301.6  &  1300.3  &  1293.1  &  1294.1  &  1277.5  &  1275.2  &  1300.8  &  1255.1  &  1224.0  \\
                   &   81.0   &   82.6   &   88.2   &   94.1   &   89.0   &   76.5   &   75.2   &   32.6   &   38.0   &   46.2   \\
 $\nu_3$           &  1326.2  &  1324.9  &  1324.7  &  1318.9  &  1319.4  &  1311.7  &  1306.1  &  1385.0  &  1355.4  &  1409.6  \\
                   &   220.2  &   217.8  &   216.5  &   210.5  &   210.6  &   221.7  &   114.6  &   218.1  &   185.5  &   53.2   \\
4$\nu_9$           &  1662.8  &  1660.4  &  1656.8  &  1656.0  &  1669.4  &  1691.0  &          &          &          &  1728.0  \\
                   &   0.727  &   0.721  &   0.641  &   0.489  &   0.787  &   67.9   &          &          &          &   0.655  \\
 $\nu_2$           &  1709.5  &  1708.1  &  1708.1  &  1702.7  &  1701.7  &  1694.2  &  1692.5  &  1770.8  &  1745.2  &  1666.0  \\
                   &   360.3  &   318.0  &   358.3  &   350.0  &   250.2  &   174.0  &   332.6  &   288.2  &   252.3  &   272.8  \\
 $\nu_1$           &  3551.6  &  3550.7  &  3551.4  &  3546.8  &  3550.2  &  3542.0  &  3542.3  &  3592.9  &  3555.5  &  3553.1  \\
                   &   83.1   &   68.3   &   83.1   &   72.6   &   79.9   &   81.6   &   85.6   &   59.6   &   89.9   &   84.8   \\
\hline\hline
\end{tabular}
\end{center}
\end{table}

\begin{table}[ht]
\scriptsize
\tabcolsep=3pt
\caption{Calculated values of the rotational energy levels (in cm$^{-1}$)
for the vibrational ground state of  H$_2$O obtained using different methods of diagonalising the Hamiltonian matrix.}\label{tab:en5}
\begin{center}
\resizebox{\columnwidth}{!}{%
\begin{tabular}{lrrrrrrrrrrrrrr}
\hline\hline
\multicolumn{3}{c|}{$J = 1$}  & \multicolumn{3}{c|}{$J = 3$}  &  \multicolumn{3}{c|}{$J = 6$}  &  \multicolumn{3}{c|}{$J = 8$}  &  \multicolumn{3}{c}{$J = 10$} \\
\hline
  Full    & Hybrid & Separate & Full & Hybrid & Separate &  Full & Hybrid & Separate &  Full & Hybrid & Separate &  Full & Hybrid & Separate \\
\hline
   23.78 &   23.78 &   23.78 &  136.48 &  136.46 &  136.63 &  445.16 &  445.15 &  446.05 &  741.25 &  741.27 &  743.30 & 1110.10 & 1110.17 & 1114.25 \\
   37.01 &   37.01 &   37.04 &  141.84 &  141.86 &  141.94 &  445.67 &  445.68 &  446.51 &  741.34 &  741.36 &  743.36 & 1110.12 & 1110.19 & 1114.25 \\
   42.33 &   42.33 &   42.34 &  173.41 &  173.38 &  173.57 &  542.57 &  542.37 &  544.34 &  881.59 &  881.57 &  885.35 & 1290.45 & 1290.80 & 1297.35 \\
         &         &         &  206.00 &  206.05 &  206.32 &  552.11 &  552.21 &  553.51 &  884.06 &  884.21 &  887.50 & 1290.99 & 1291.38 & 1297.73 \\
         &         &         &  212.04 &  212.07 &  212.38 &  603.25 &  603.00 &  605.72 &  983.49 &  983.38 &  989.97 & 1437.91 & 1438.88 & 1449.77 \\
         &         &         &  285.02 &  284.97 &  286.59 &  648.63 &  649.02 &  651.08 & 1005.67 & 1006.26 & 1011.01 & 1445.39 & 1446.60 & 1456.00 \\
         &         &         &  285.23 &  285.18 &  286.80 &  661.79 &  662.03 &  664.64 & 1051.46 & 1051.91 & 1059.19 & 1540.49 & 1542.89 & 1557.39 \\
         &         &         &         &         &         &  756.87 &  757.49 &  761.98 & 1122.96 & 1124.40 & 1130.72 & 1582.01 & 1584.84 & 1596.30 \\
         &         &         &         &         &         &  757.99 &  758.61 &  763.12 & 1132.61 & 1134.11 & 1141.04 & 1619.15 & 1623.20 & 1637.39 \\
         &         &         &         &         &         &  890.14 &  890.12 &  901.68 & 1256.47 & 1258.37 & 1269.40 & 1720.48 & 1724.91 & 1739.20 \\
         &         &         &         &         &         &  890.18 &  890.15 &  901.71 & 1257.27 & 1259.21 & 1270.23 & 1726.95 & 1731.94 & 1746.51 \\
         &         &         &         &         &         & 1049.72 & 1046.94 & 1073.55 & 1415.37 & 1416.12 & 1439.06 & 1878.59 & 1883.34 & 1906.06 \\
         &         &         &         &         &         & 1049.72 & 1046.94 & 1073.55 & 1415.40 & 1416.16 & 1439.09 & 1879.12 & 1883.97 & 1906.63 \\
         &         &         &         &         &         &         &         &         & 1599.11 & 1595.54 & 1641.45 & 2061.68 & 2064.15 & 2105.58 \\
         &         &         &         &         &         &         &         &         & 1599.11 & 1595.54 & 1641.45 & 2061.70 & 2064.19 & 2105.60 \\
         &         &         &         &         &         &         &         &         & 1805.18 & 1791.73 & 1876.13 & 2267.96 & 2263.69 & 2338.32 \\
         &         &         &         &         &         &         &         &         & 1805.18 & 1791.73 & 1876.13 & 2267.96 & 2263.69 & 2338.32 \\
         &         &         &         &         &         &         &         &         &         &         &         & 2494.83 & 2476.44 & 2603.63 \\
         &         &         &         &         &         &         &         &         &         &         &         & 2494.83 & 2476.44 & 2603.63 \\
         &         &         &         &         &         &         &         &         &         &         &         & 2740.26 & 2696.06 & 2901.09 \\
         &         &         &         &         &         &         &         &         &         &         &         & 2740.26 & 2696.06 & 2901.09 \\
\hline\hline
\end{tabular}
}
\end{center}
\end{table}

\begin{table}[ht]
\scriptsize
\tabcolsep=3pt
\caption{Calculated values of the rotational energy levels (in cm$^{-1}$)
for the vibrational ground state of HNOH$_3$  obtained using different methods of diagonalizing the Hamiltonian matrix.}\label{tab:en6}
\begin{center}
\resizebox{\columnwidth}{!}{%
\begin{tabular}{lrrrrrrrrrrrrrr}
\hline\hline
\multicolumn{3}{c|}{$J = 1$}  & \multicolumn{3}{c|}{$J = 10$}  &  \multicolumn{3}{c|}{$J = 15$}  &  \multicolumn{3}{c|}{$J = 45$}  &  \multicolumn{3}{c}{$J = 60$} \\
\hline
  Full    & Hybrid & Separate & Full & Hybrid & Separate &  Full & Hybrid & Separate &  Full & Hybrid & Separate &  Full & Hybrid & Separate \\
\hline
    0.61 &    0.61 &    0.61 &   53.18 &   53.18 &   53.18 &  200.17 &  200.17 &  200.21 &  440.89 &  440.89 &  441.09 &  775.20 &  775.19 &  775.81 \\
    0.64 &    0.64 &    0.64 &   59.24 &   59.24 &   59.24 &  212.49 &  212.49 &  212.54 &  477.62 &  477.62 &  477.86 &  824.40 &  824.40 &  825.12 \\
    0.84 &    0.84 &    0.84 &   64.88 &   64.88 &   64.88 &  224.39 &  224.39 &  224.45 &  512.65 &  512.65 &  512.96 &  871.90 &  871.90 &  872.76 \\
         &         &         &   70.09 &   70.09 &   70.10 &  246.93 &  246.93 &  247.01 &  545.98 &  545.98 &  546.38 &  917.69 &  917.69 &  918.72 \\
         &         &         &   74.89 &   74.89 &   74.90 &  257.56 &  257.56 &  257.66 &  577.63 &  577.63 &  578.12 &  961.77 &  961.77 &  963.00 \\
         &         &         &   79.27 &   79.27 &   79.28 &  267.78 &  267.78 &  267.89 &  607.58 &  607.58 &  608.18 & 1004.15 & 1004.14 & 1005.60 \\
         &         &         &   83.22 &   83.22 &   83.23 &  286.94 &  286.94 &  287.09 &  635.84 &  635.84 &  636.56 & 1044.82 & 1044.82 & 1046.52 \\
         &         &         &   86.74 &   86.74 &   86.76 &  295.89 &  295.89 &  296.06 &  662.41 &  662.41 &  663.25 & 1083.80 & 1083.80 & 1085.76 \\
         &         &         &   89.83 &   89.83 &   89.85 &  304.42 &  304.42 &  304.60 &  687.29 &  687.28 &  688.26 & 1121.08 & 1121.08 & 1123.32 \\
         &         &         &   92.48 &   92.48 &   92.50 &  320.20 &  320.20 &  320.42 &  710.47 &  710.47 &  711.57 & 1156.67 & 1156.66 & 1159.19 \\
         &         &         &   94.62 &   94.62 &   94.64 &  327.45 &  327.45 &  327.69 &  731.95 &  731.94 &  733.18 & 1190.56 & 1190.55 & 1193.38 \\
         &         &         &   96.09 &   96.09 &   96.12 &  334.28 &  334.28 &  334.54 &  751.73 &  751.71 &  753.08 & 1222.76 & 1222.74 & 1225.87 \\
         &         &         &   97.16 &   97.16 &   97.19 &  346.64 &  346.64 &  346.94 &  769.77 &  769.76 &  771.25 & 1253.26 & 1253.24 & 1256.67 \\
         &         &         &   98.54 &   98.55 &   98.57 &  352.17 &  352.16 &  352.48 &  786.07 &  786.05 &  787.67 & 1282.06 & 1282.03 & 1285.77 \\
         &         &         &  100.47 &  100.47 &  100.49 &  357.25 &  357.24 &  357.58 &  800.58 &  800.53 &  802.29 & 1309.15 & 1309.11 & 1313.16 \\
         &         &         &  102.76 &  102.75 &  102.78 &  366.00 &  365.98 &  366.37 &  813.21 &  813.13 &  815.03 & 1334.53 & 1334.48 & 1338.84 \\
         &         &         &         &         &         &  369.57 &  369.52 &  369.96 &  823.70 &  823.48 &  825.65 & 1358.19 & 1358.12 & 1362.78 \\
         &         &         &         &         &         &  372.34 &  372.25 &  372.76 &  830.85 &  830.33 &  832.97 & 1380.09 & 1380.01 & 1384.97 \\
         &         &         &         &         &         &  376.42 &  376.42 &  376.83 &  837.09 &  837.27 &  839.13 & 1400.23 & 1400.12 & 1405.39 \\
         &         &         &         &         &         &  379.27 &  379.36 &  379.67 &  846.55 &  847.12 &  848.51 & 1418.54 & 1418.38 & 1423.97 \\
         &         &         &         &         &         &  382.75 &  382.88 &  383.15 &  857.83 &  858.41 &  859.73 & 1434.96 & 1434.71 & 1440.65 \\
         &         &         &         &         &         &  390.87 &  390.93 &  391.24 &  870.48 &  870.69 &  872.29 & 1449.30 & 1448.86 & 1455.28 \\
         &         &         &         &         &         &  395.40 &  395.35 &  395.76 &  884.34 &  883.72 &  886.04 & 1460.97 & 1459.67 & 1467.32 \\
         &         &         &         &         &         &  400.22 &  400.00 &  400.55 &         &         &         & 1468.58 & 1467.55 & 1475.24 \\
         &         &         &         &         &         &         &         &         &         &         &         & 1490.44 & 1492.11 & 1496.62 \\
         &         &         &         &         &         &         &         &         &         &         &         & 1520.79 & 1522.31 & 1526.63 \\
         &         &         &         &         &         &         &         &         &         &         &         & 1556.46 & 1555.09 & 1561.82 \\
\hline\hline
\end{tabular}
}
\end{center}
\end{table}

\appendix
\newpage
\section {The Hamiltonian}

Our rotation-vibration Hamiltonian written in curvilinear internal coordinates and an
Eckart embedding has the form \cite{88GrPa.method}
\begin{equation}
\hat H_{vr} = \hat H_v
- \frac{\hbar^2}{2} \sum_{a, b} \frac{\partial}{\partial \xi_a}
\mu_{ab} (\underline{q}) \frac{\partial}{\partial \xi_b}
~, ~ ~ ~ ~ ~ ~ \xi_a, \xi_b = \alpha, \beta, \gamma ~,
\end{equation}
where $\underline{\xi}$ are the rotational coordinates and
$\hat H_v$ is the vibrational part of the Hamiltonian
\begin{equation}
\hat H_v = \hat T_v + V (\underline{q})
\end{equation}
\begin{equation}
\hat T_v =
- \frac{\hbar^2}{2} \sum_{i,j} t^{\frac{1}{4}} \frac{\partial}{\partial q_i}
                  g_{ij}(\underline{q}) t^{-\frac{1}{2}} \frac{\partial}{\partial q_j} t^{\frac{1}{4}}.
\end{equation}
Here $q_i$ are internal, vibrational curvilinear coordinates given by
changes in bond lengths or changes in the valence bond angles,
dihedral angles, etc.; $\alpha$, $\beta$, $\gamma$ are the Euler
angles between the axes of the equilibrium moment of inertia tensor
and external Cartesian coordinate axes; $\mu_{ab}(\underline{q})$ are
elements of the inverse of the moment of inertia tensor,
$\underline{I}(\underline{q})$; $\hat T_v$ is the vibrational kinetic
energy operator and $g_{ij}(\underline{q})$ are elements of the
kinetic energy coefficients matrix $\underline{G}(\underline{q})$ and
$t = \det[\underline{G}]$. Finally, $V(\underline{q})$ is the
molecular potential energy.

After transformation, the vibrational kinetic operator can be written as
\begin{equation}
\hat T_v =
- \frac{\hbar^2}{2} \sum_{i,j} \frac{\partial}{\partial q_i}
                  g_{ij}(\underline{q}) \frac{\partial}{\partial q_j}
+ \beta(\underline{q}),
\end{equation}
where
\begin{equation}
\begin{split}
{\beta(\underline{q})}&= -\frac{\hbar^2}{2} \sum_{i,j}{\left\{\frac{{\partial}{g}_{ij}(\underline{q})}{{\partial}{q_i}}
\sum_{k,l}{\zeta}_{kl}(\underline{q})\frac{{\partial}{g}_{kl}(\underline{q})}{{\partial}{q_j}}
\right.}+\\
+\frac{1}{4}{g}_{ij}(\underline{q})&\Bigl[\sum_{k,l}{\zeta}_{kl}(\underline{q})
\frac{{{\partial}^2}{g}_{kl}(\underline{q})}{{\partial}{q_i}{\partial}{q_j}}-
\sum_{k,l,m,n}{\zeta}_{kl}(\underline{q}){\zeta}_{mn}(\underline{q})\times\Bigr.\\
&\times\left.\Bigl.\Bigl(\frac{{\partial}{g}_{lm}(\underline{q})}{{\partial}{q_i}}
\frac{{\partial}{{g}_{kn}(\underline{q})}}{{\partial}{q_j}}+
\frac{{\partial}{{g}_{kl}(\underline{q})}}{{\partial}{q_i}}\frac{{\partial}
{g}_{mn}(\underline{q})}{{\partial}{q_j}}\Bigr)\Bigr]\right\},
\end{split}
\end{equation}
is the so-called pseudo-potential \cite{watsonterm}; $\zeta_{ij}(\underline{q})$  are
elements of $\underline{G}(\underline{q})^{-1}$.

$\underline{G}(\underline{q})$ is a complicated function of the internal coordinates,
with the elements in general case  given elsewhere \cite{88GrPa.method}. If the
coordinates associated with both indices $i$ and $j$ represent changes in the bond
lengths then
\begin{equation}
g_{ij}(\underline{q}) = g_{ij}^{0}(\underline{\varphi})~;
\end{equation}
if the coordinate $i$ represents a change in the bond length and $j$ is an angular
coordinate then
\begin{equation}
g_{ij}(\underline{q}) = \sum_{k} \frac{1}{r_k} g_{ij}^{k}(\underline{\varphi})~,
\end{equation}
and if $i$ and $j$ both represent angular coordinates it becomes
\begin{equation}
g_{ij}(\underline{q}) = \sum_{k,l} \frac{1}{r_k r_l} g_{ij}^{kl}(\underline{\varphi}) ~.
\end{equation}
In these expressions $r_k$ is bondlength of the $k$-th bond and $\underline{\varphi}$
represents the angular coordinates.

The computation of $\underline{G}(\underline{q})$
is achieved using a second-order Taylor expansion in the angular coordinates
\begin{equation}
g_{ij}(\underline{q}) = g_{ij}^{0}(0)
     + \sum_{m} \left( \frac{\partial g_{ij}^{0}(\underline{\varphi})}{\partial \varphi_m} \right)_0 \varphi_m
     + \frac{1}{2} \sum_{m,n}
       \left( \frac{\partial^2 g_{ij}^{0}(\underline{\varphi})}{\partial \varphi_m \partial \varphi_n} \right)_0 \varphi_m \varphi_n ~~~,
\end{equation}
\begin{equation}
g_{ij}(\underline{q}) = \sum_{k} \frac{1}{r_k}
       \left[ g_{ij}^{k}(0)
     + \sum_{m} \left( \frac{\partial g_{ij}^{k}(\underline{\varphi})}{\partial \varphi_m} \right)_0 \varphi_m
     + \frac{1}{2} \sum_{m,n}
       \left( \frac{\partial^2 g_{ij}^{k}(\underline{\varphi})}{\partial \varphi_m \partial \varphi_n} \right)_0 \varphi_m \varphi_n \right] ~~~,
\end{equation}
\begin{equation}
g_{ij}(\underline{q}) = \sum_{k,l} \frac{1}{r_k r_l}
       \left[ g_{ij}^{kl}(0)
     + \sum_{m} \left( \frac{\partial g_{ij}^{kl}(\underline{\varphi})}{\partial \varphi_m} \right)_0 \varphi_m
     + \frac{1}{2} \sum_{m,n}
       \left( \frac{\partial^2 g_{ij}^{kl}(\underline{\varphi})}{\partial \varphi_m \partial \varphi_n} \right)_0 \varphi_m \varphi_n \right] ~.
\end{equation}
These formulae are exact as $g_{ij}^{0}(\underline{\varphi})$,
$g_{ij}^{k}(\underline{\varphi})$ and
$g_{ij}^{kl}(\underline{\varphi})$ are quadratic functions of angles
when the angular coordinates.  The
$\underline{\varphi}$ are represented
 as cosine differences $\cos \varphi_i - \cos
\varphi_i^e$, where $\cos \varphi_i^e$ is the instantaneous
equilibrium angle for bond angles, and sine differences for  dihedral angles.

To calculate the vibrational Hamiltonian matrix elements in block 2, which give
the perturbative contribution to block 1,  the vibrational kinetic energy
coefficients are expanded in the polynomial form and truncated after the second
order
\begin{equation}
g_{ij}(\underline{q}) = g_{ij}(0)
     + \sum_{m} \left( \frac{\partial g_{ij}(\underline{q})}{\partial q_m} \right)_0 q_m
     + \frac{1}{2} \sum_{m,n}
       \left( \frac{\partial^2 g_{ij}(\underline{q})}{\partial q_m \partial q_n} \right)_0 q_m q_n ~.
\end{equation}
This form is convenient because it allows faster (by an order-of-magnitude or more)
computation of the coefficients without significant loss of accuracy.

The potential energy function used by us is a fourth-order polynomial
\begin{equation}
V(\underline{q}) = \frac{1}{2}  \sum_{i,j} D_{ij} x_i x_j
     + \frac{1}{6}  \sum_{i,j,k} D_{ijk} x_i x_j x_k
     + \frac{1}{24} \sum_{i,j,k,l} D_{ijkl} x_i x_j x_k x_l ~,
\end{equation}
where a Morse transformation, $x_i = (1-\exp^{-\alpha_i \Delta r_i})$, is used to
represent  changes in terminal bonds such as  X--H and $x_i = q_i$ for angular
coordinates and the coordinates of skeletal changes in the bond lengths.

In variational calculations of  vibrational energy
levels, the Hamiltonian matrix elements
\begin{equation}
H_{kn,ml} = \langle \chi_{kn} \arrowvert \hat H_v \arrowvert \chi_{ml} \rangle ~~~
\end{equation}
are computed using the product form
\begin{equation}
\chi_{kn} = \prod_{i} \phi_{k_{i}}(r_i) \prod_{s} \psi_{n_{s}}(Q_s)
\end{equation}
of the basis functions, which are eigenfunctions of the Morse or harmonic oscillators.
Morse oscillator functions, $\phi_{k_{i}} (r_i)$, are used for
 the stretching coordinates, $r_i$, for which the potential is given using a Morse
transformation. Harmonic basis functions, $\psi_{n_{s}} (Q_s)$, are used for
the other coordinates which are represented using the curvilinear normal
coordinates
\begin{equation}
Q_s = \sum_{i} L^{(q)}_{is} q_i
\end{equation}
expressed as a linear sum over the internal coordinates, $q_i$, for which the
potential function is  defined as a Taylor series. The coordinates $Q_s$ are
those which diagonalize the harmonic part of the Hamiltonian given in the
internal coordinates $q_i$. With these definitions, all multidimensional
integrals required to calculate the Hamiltonian matrix elements  are separated
into products of one-dimensional integrals between either Morse functions or
harmonic oscillators. All these integrals have a simple analytic form which
results in high-speed computation of the Hamiltonian matrix elements.
Diagonalizing the Hamiltonian matrix gives the anharmonic vibrational energy
levels $E^{vib}_\lambda$ and the corresponding wave functions
$\phi^{vib}_\lambda$.

For ro-vibrational energy levels it is necessary to calculate elements of the
complex Hermitian Hamiltonian matrix
\begin{equation}
H^{JJ\p}_{\lambda k m, \lambda\p k\p m\p} =
\langle \chi^{J}_{\lambda k m} \arrowvert \hat H_{vr} \arrowvert
\chi^{*J\p}_{\lambda\p k\p m\p} \rangle ~,
\end{equation}
using the variational basis functions
\begin{equation}
\chi^{J}_{\lambda km} = \Phi^{vib}_{\lambda} \phi^{J}_{km} ~~~,~~~
\phi^{J}_{km} = \left( \frac{2J+1}{8 \pi^2} \right)^{1/2} D^{J~*}_{km},
\end{equation}
where $D^{J}_{km}$ is a (complex) Wigner function. In this case
\begin{equation}
H^{JJ\p}_{\lambda k m, \lambda\p k\p m\p} =
E^{vib}_\lambda \delta_{\lambda \lambda\p} \delta_{JJ\p} \delta_{kk\p} \delta_{mm\p} -
\langle \Phi^{vib}_{\lambda} \phi^{J}_{k m} \arrowvert
\frac{\hbar^2}{2} \sum_{a,b} \frac{\partial}{\partial \xi_a}
                   \mu_{ab}(\underline{q}) \frac{\partial}{\partial \xi_b}
\arrowvert \Phi^{vib}_{\lambda\p} \phi^{*J\p}_{k\p m\p} \rangle
\end{equation}
This expression is equivalent to
\begin{equation}
H^{JJ\p}_{\lambda k m, \lambda\p k\p m\p} = H^{J}_{\lambda k, \lambda\p k\p } \delta_{J,J\p}  \delta_{mm\p},
\end{equation}
where
\begin{equation}
H^{J}_{\lambda k, \lambda\p k\p } =
E^{vib}_\lambda \delta_{\lambda \lambda\p} \delta_{JJ\p} \delta_{kk\p} \delta_{mm\p}
- \frac{\hbar^2}{2} \sum_{a,b} \bar{\mu}^{\lambda \lambda\p}_{ab}
  \langle \phi^{J}_{km} \arrowvert \frac{\partial^2}{\partial \xi_a \partial \xi_b}
  \arrowvert \phi^{*J}_{k\p m\p} \rangle \delta_{JJ\p} \delta_{mm\p},
\end{equation}
and
\begin{equation}
\bar{\mu}^{\lambda \lambda\p}_{ab} =
\langle \Phi^{vib}_{\lambda} \arrowvert \mu_{ab}(\underline{q}) \arrowvert \Phi^{vib}_{\lambda\p} \rangle.
\end{equation}
To calculate $\bar{\mu}^{\lambda \lambda\p}_{ab}$, the matrix elements
$\mu_{ab}(\underline{q})$ are expanded to the second-order as a Taylor series
\begin{equation}
\mu_{ab}(\underline{q}) = \mu_{ab}(0)
     + \sum_{m} \left( \frac{\partial \mu_{ab}(\underline{q})}{\partial q_m} \right)_0 q_m
     + \frac{1}{2} \sum_{m,n}
       \left( \frac{\partial^2 \mu_{ab}(\underline{q})}{\partial q_m \partial q_n} \right)_0 q_m q_n,
\end{equation}
where
\begin{equation*}
\left(\frac{\partial\mu_{ab}(\underline{q})}{\partial q_m}\right)_0 =
-\sum_{i,j}\frac{\left(\frac{\partial I_{ij}(\underline{q})}{\partial q_m}\right)_0}{I_{ia}(0) I_{ib}(0)},
\end{equation*}
\begin{equation}
\left(\frac{\partial^2\mu_{ab}(\underline{q})}{\partial q_m\partial q_n}\right)_0 =
-\sum_{i,j}\frac{\left(\frac{\partial^2 I_{ij}(\underline{q})}{\partial q_m\partial q_n}\right)_0}{I_{ia}(0) I_{ib}(0)}
+\sum_{i,j,k,l}\left[\frac{\left(\frac{\partial I_{ij}(\underline{q})}{\partial q_m}\right)_0
\left(\frac{\partial I_{kl}(\underline{q})}{\partial q_n}\right)_0}{I_{ia}(0) I_{jk}(0) I_{lb}(0)}
+ \frac{\left(\frac{\partial I_{ij}(\underline{q})}{\partial q_m}\right)_0
\left(\frac{\partial I_{kl}(\underline{q})}{\partial q_n}\right)_0}{I_{jb}(0) I_{ik}(0) I_{la}(0)} \right] ~.
\end{equation}
In this case
\begin{equation*}
\bar{\mu}^{\lambda \lambda\p}_{ab} =
\langle \Phi^{vib}_{\lambda} \arrowvert \mu_{ab}(\underline{q})
\arrowvert \Phi^{vib}_{\lambda\p} \rangle =
\end{equation*}
\begin{equation}
\mu_{ab}(0) \delta_{\lambda \lambda\p} +
\sum_{m} \left( \frac{\partial \mu_{ab}(\underline{q})}{\partial q_m} \right)_0
    \langle \Phi^{vib}_{\lambda} \arrowvert  q_m \arrowvert \Phi^{vib}_{\lambda\p} \rangle +
\frac{1}{2} \sum_{m,n}
  \left( \frac{\partial^2 \mu_{ab}(\underline{q})}{\partial q_m \partial q_n} \right)_0
    \langle \Phi^{vib}_{\lambda} \arrowvert q_m q_n \arrowvert \Phi^{vib}_{\lambda\p} \rangle
\end{equation}
and all the integrals reduce to products of one-dimensional integrals over
either Morse or harmonic oscillators.

When analyzing the off-diagonal elements
of the vibration-rotation Hamiltonian matrix
\begin{equation}
H^{J}_{\lambda k, \lambda\p k\p} =
- \frac{\hbar^2}{2} \sum_{a,b} \bar{\mu}^{\lambda \lambda\p}_{ab}
  \langle \phi^{J}_{km} \arrowvert \frac{\partial^2}{\partial \xi_a \partial \xi_b}
  \arrowvert \phi^{*J}_{k\p m\p} \rangle \delta_{JJ\p} \delta_{mm\p}~,
\end{equation}
it can be seen that they differ significantly in magnitude, depending on
whether they are diagonal in the vibrations, $\lambda = \lambda\p$, or couple
different vibrational states,  $\lambda \ne \lambda\p$. For semi-rigid
molecules with small values of the vibrational quantum numbers the following
condition usually holds
\begin{equation}
|H^{J}_{\lambda k, \lambda k\p }| > |H^{J}_{\lambda k, \lambda\p k\p }| ~~~,~~~ \lambda \ne \lambda\p.
\end{equation}
This results from the slight change in the effective geometry of the molecule upon
vibrational excitation
\begin{equation}
|\mu_{ab}(0)| >
|\sum_{m} \left( \frac{\partial \mu_{ab}(\underline{q})}{\partial q_m} \right)_0
    \langle \Phi^{vib}_{\lambda} \arrowvert  q_m \arrowvert \Phi^{vib}_{\lambda\p} \rangle +
\frac{1}{2} \sum_{m,n}
  \left( \frac{\partial^2 \mu_{ab}(\underline{q})}{\partial q_m \partial q_n} \right)_0
    \langle \Phi^{vib}_{\lambda} \arrowvert q_m q_n \arrowvert \Phi^{vib}_{\lambda\p} \rangle|
\end{equation}
and
\begin{equation}
|\bar{\mu}^{\lambda \lambda}_{ab}| > |\bar{\mu}^{\lambda \lambda\p}_{ab}|  ~,~~~ \lambda \ne \lambda\p.
\end{equation}

When calculating the vibrational-rotational energy levels, the off-diagonal elements
$H^{J}_{\lambda k, \lambda\p k\p}$ corresponding to different vibrational states
$\lambda \ne \lambda\p$ give a much smaller contribution (change in the diagonal elements
in the block that will be diagonalized) to the calculated energy levels than the non-diagonal elements
$H^{J}_{\lambda k, \lambda k\p}$ within the vibrational state in question.
These changes are given approximately by
\begin{equation}
\Delta E_{\lambda \lambda\p} = \frac{\left({H^{J}_{\lambda k, \lambda\p k\p}}\right)^{2}}
{H^{J}_{\lambda k, \lambda k} - H^{J}_{\lambda\p k\p, \lambda\p k\p}},
\end{equation}
\begin{equation}
\Delta E_{\lambda \lambda} = \frac{\left({H^{J}_{\lambda k, \lambda k\p}}\right)^{2}}
{H^{J}_{\lambda k, \lambda k} - H^{J}_{\lambda k\p, \lambda k\p}} ~.
\end{equation}
However,
\begin{equation}
|H^{J}_{\lambda k, \lambda k} - H^{J}_{\lambda\p k\p, \lambda\p k\p}| \gg
|H^{J}_{\lambda k, \lambda k} - H^{J}_{\lambda k\p, \lambda k\p}|
\end{equation}
since $(H^{J}_{\lambda k, \lambda k} - H^{J}_{\lambda k\p, \lambda k\p})$ involves only
a change in the rotational energy level, while
$(H^{J}_{\lambda k, \lambda k} - H^{J}_{\lambda\p k\p, \lambda\p k\p})$
involves also a change in the vibrational energy level.

\end{document}